\documentclass[AMA,STIX1COL]{WileyNJD-v2}
\usepackage{subcaption}
\articletype{Research Article}%

\received{26 April 2016}
\revised{6 June 2016}
\accepted{6 June 2016}

\begin{document}


\title{SOS: Socially Omitting Selfishness in IoT for Smart and Connected Communities}

\author[1,2]{Ghani ur Rehman}

\author[1]{Anwar Ghani*}

\author[2]{Muhammad Zubair}

\author[1]{Muhammad Imran Saeed}

\author[3]{Dhananjay Singh (Senior Member, IEEE)}



\authormark{Ghani \textsc{et al}}

\address[1]{\orgdiv{Department of Computer Science \& Software Engineering}, \orgname{International Islamic University}, \orgaddress{\state{Islamabad 44000}, \country{Pakistan}}}

\address[2]{\orgdiv{Faculty of Computer Science \& Bioinformatics}, \orgname{Khushal Khan Khattak University}, \orgaddress{\state{Karak 27000}, \country{Pakistan}}}

\address[3]{Department of Electronics Engineering, \orgname{Hankuk University of Foreign Studies}, \orgaddress{\state{Yongin 449-791}, \country{South Korea}}}




\corres{*Dr. Anwar Ghani, Department of Computer Science \& Software Engineering, International Islamic University Islamabad, \email{anwar.ghani@iiu.edu.pk}}

\presentaddress{Department of Computer Science \& Software Engineering, International Islamic University Islamabad}

\abstract[Summary]{Smart and Connected Communities (SCC) is an emerging field of Internet of Things (IoT), and it is having potential applications to improve human life. The improvement may be in terms of preservation, revitalization, livability, and sustainability of a community. The resources of the nodes and devices in the SCC have certain constraints that may not allow the devices and nodes to cooperate to save their resources such as memory, energy, and buffer, or simply maximize their performance. Thus, to stimulate the nodes to avoid selfish behavior, SSC needs a novel and well-organized solution to motivate nodes for cooperation. This article aims to resolve the issue of selfish behaviors in SCC and to encourage the nodes for cooperation. A novel mechanism Socially Omitting Selfishness (SOS) has been proposed to manage/eradicate selfishness using a socially-oriented election process. The election process elects different heads based on weight and cooperation (using VCG model). The election of heads and incentive mechanism encourages the nodes to show participation and behave as highly cooperative members of the community. Furthermore, an extended version of the Dempster-Shafer model has been used to discourage the selfish behavior of the participating nodes in the SOS scheme. It uses different monitoring and gateway nodes to efficiently employ the proposed scheme. A mathematical model has been developed for the aforementioned aspects and simulated through NS2 simulation environment to analyze the performance of SOS. The results of the proposed scheme outperform the contemporary schemes in terms of average delivery delay, packet delivery ratio, throughput, and average energy.}

\keywords{Sustainability, IoT, Revitalization, Smart and connected communities, Social selfishness, Selfish behaviors,Incentive techniques}

\jnlcitation{\cname{%
\author{G. Rahman}, 
\author{A. Ghani}, 
\author{M. Zubair}, 
\author{M. I. Saeed}, and
\author{D. Singh}} (\cyear{2019}), 
\ctitle{SOS: Socially Omitting Selfishness in IoT for Smart and Connected Communities}, \cjournal{International Journal of ABC, XYZ}, \cvol{2019;00:1--16}
}

\maketitle

\footnotetext{\textbf{Abbreviations:} SOS, Socially Omiting selfishness; IoT, Internet of Things; SCC, Smart and Connected Communities}

\section{INTRODUCTION}
\label{intro}

Internet of things (IoT) is an emerging area in modern communication networks. It consists of different devices (things) such as machines, people, animals, and objects embedded with sensors and actuators~\cite{01}. Data sensed by these sensors and actuators are relayed to remote servers for further processing. The processing time and storage capacity depends on the capabilities of IoT objects~\cite{02,03,04,05,06,07,08}. The advancements in IoT are influential on the smart and connected Communities (SCC). The main considerations of the SCC are the present past and future of the major areas of society~\cite{09,10}. Aims and objectives of SCC are the preservation, livability, revitalization, and sustainability of human life to make life easy by remembering the past of human life, focus on present, and plan for a better future. The cultural heritage for communities is preserved and living needs are referred to as livability~\cite{11}. Sustainability is the need of making plans for the future having three important pillars such as social, environmental, and economic aspects~\cite{12}. The nodes in SCC can store and forward data to the destination hop by hop. It means nodes need to be socially cooperative to relay data. However, some of the nodes adopt a selfish nature to save their resources. 

Limited resources, preservation of bandwidth, manipulation, energy savings and other social behavior of the nodes make the node self-centered and selfish. Selfishness may be categorized as individual and social~\cite{13,14,14a}. The individual selfishness has the same degree of selfishness with all other nodes that may not forward messages even to its neighbor. Social selfishness has a different degree of selfishness because the relationship varies with all other nodes in the network. The nodes having a variable relationship with neighbor nodes may forward messages to their friends, and refuse to forward messages to the strangers. The selfish behavior of the node has an adverse effect on the throughput of the network and can affect the performance of the network by losing connectivity~\cite{15}. In this article, the problem of selfishness either individual or social has been considered to design a scheme that can omit the selfishness using a social mechanism of election in SCC.

There are many existing techniques such as the Game-theoretic reward-based system and the Cooperative Watchdog System that addressed the issue of selfishness and motivate node for cooperation. Different types of cards are assigned to nodes according to their selfishness level in the first scheme. In the second scheme, a reputation score is assigned to each node. However, these schemes have issues like (i) Individual importance factor problem: in~\cite{27}, only individual importance of a node is considered for trust. (ii) monitoring nodes can be selfish: this scheme only monitors the behaviors of the selfish node. But monitoring node can also be selfish as they give wrong information about the selfish or cooperative nodes (iii) strict punishment issue: the node can not join the network once it has been punished~\cite{28}. Node Selfishness is omitted differently in proposed SOS scheme. It also monitors the behavior of the monitoring nodes in the network that is ignored in the above two schemes. Another feature of the proposed scheme is the computation of the collective trust of monitoring nodes. The collective trust called Collective Importance Factor (CIF) is computed to entitled any monitoring node to be cooperative or selfish. It also gives a chance to the selfish node after showing selfishness for the first time. A participating node can rejoin the network by making negative payment first. At last, the node can be  expelled from the network for showing selfish behavior repeatedly.

In this article, a new motivation system is presented called socially omitting selfishness. In this system, selfish nodes are encouraged to take part in the election. SOS communities are headed by a different head, elected in the election process. The different types of elected heads are community head ($CH$), monitoring head ($MH$), and incentive head ($IH$). These community heads are elected based on certain features such as  number of votes they get in election, weight, and cooperation. Each of these heads has their own duties in the community. The nodes that have malicious or selfish behaviors are fined. SOS scheme motivates the selfish nodes for cooperation in SCC. The primary contributions of the proposed SOS scheme in SCC are:
\begin{itemize}
\item To analyzed the selfish behavior of nodes in SCC both individually and socially. Furthermore, to analyze the effect of nodes participating in the overall performance of the network.
\item To design a novel strategy that motivates the non-cooperative nodes for participation within the community to improve network performance in SCC.
\item To formulate criteria taking into account certain parameters like weight and cooperation that can be used to elect various heads such as Community Head, Monitoring Head, and Incentive Head.
\item To develop a mechanism for solving the weight tie issue by adding cooperation as a parameter for the next nomination criteria in the election of heads in the election process.
\item To introduce an effective monitoring scheme that constantly monitors the selfish behavior of the nodes to calculate the Collective Importance Factor (CIF) and track the behavior of the monitoring head as well to avoid any discriminatory behavior on the part of the monitoring head itself.
\item To perform comparative analysis of the proposed scheme, different incentive-based schemes are compared with the proposed scheme to gauge the performance of the proposed scheme in terms of improving human livability, revitalization, improved packet delivery ratio, average delivery delay, and energy constraint.
\end{itemize}

The remaining of the paper is organized in the following sections: Section~\ref{litrev} discusses related works. Section~\ref{sys} shows the detailed design of SOS. Section~\ref{per} shows the performance evaluation of the proposed SOS scheme. The paper is concluded and future work is discussed in section~\ref{con}.
\section{RELATED WORK}
\label{litrev}
SCC has the number of nodes that are selfish and non-cooperative in nature. The selfish behavior of the nodes has been widely examined and is highly interested in the researchers. Selfish nodes are degrading the overall performance of the network~\cite{16}. The incentive-based techniques are implemented to encourage the selfish nodes for cooperation and share its resources altruistically with other nodes in the network~\cite{17},~\cite{17a}. The incentive-based technique is further classified into four classes such as reputation-based, game-theoretic-based, credit-based and barter-based system.

The reputation-based motivation system depends on the degree of node cooperation within the network. Compared to non-cooperative nodes, cooperative nodes are extremely appreciated. The credit-based scheme operates to give the nodes some benefits for showing cooperation. The nodes can subsequently utilize this awarded credit for their purpose later. Barter-based motivation systems, also known as the Tit-For-Tat (TFT) approach, where nodes share the same information.

Yuxin et al.~\cite{18} proposed a game-based incentive scheme. Two types of relationship such as competitive and cooperative are considered. To motivate the nodes for cooperation, a contribution measurement is given during the game. Annalisa et al.~\cite{19} proposed a social scheme called SORSI for detecting selfishness and encourage the selfish nodes for data forwarding. Ning et al.~\cite{20} proposed a CAIS scheme to discourage selfishness in social networks. In this scheme, social interactions among nodes make communities in the network. The nodes are rewarded two types of credit namely social credit and non-social credit for data forwarding within the same communities or different communities.

Wang et al.~\cite{21} proposed a hop limited flooding scheme to tackle the issue of selfishness in Delay Tolerant Networks. Jedari et al.~\cite{22} proposed a social-based watchdog system to detect selfish nodes in an opportunistic mobile network. This scheme differentiates the degree of the selfishness of the nodes because the punishment and rewarding process employed to encourage nodes might not be the same for all nodes in the network. Wang et al.~\cite{23} presented an incentive approach to resolve the issue of selfishness in the urban environment. Seregina et al.~\cite{24} addressed the issue of selfishness by proposes a reward-based mechanism to motivate the mobile nodes for cooperation.

Lu et al.~\cite{25} proposed a Geographic information and node selfish-based scheme to tackle the problem of selfishness. To choose a decent next-hop relay node, the readiness of the relay node is merged with geographic informations. Wei et al.~\cite{26} proposed a community-based and reputation-based incentive scheme to motivate a selfish node to take part in data forwarding. In this scheme each node can retain, update and show reputation for verification whenever necessary. The critical factor in this scheme is the altruism function that kicks out selfish nodes.

In the research article~\cite{27}, the author presents a game theoretical reward-based system to manage selfish nodes in the network. Dias et al.~\cite{28} proposed a cooperative watchdog system to tackle the issue of selfishness by assigning a reputation score to all nodes in the network. Fawaz et al.~\cite{29} proposed an incentive mechanism to motivate the selfish nodes for cooperation in the vehicular networks. In the research article~\cite{30}, the author proposed a barter-based scheme to resolve the issue of selfishness in the network.  Liu et al.~\cite{31} also proposed the barter-based scheme among different communities to encourage selfish nodes for taking part in message forwarding.

Li et al.~\cite{32} proposed a Social scheme to manage selfishness in the network.  In a social community or group, the nodes with strong social interactions with one another are likely willing to forward data. In the research article~\cite{33}, the author proposed an incentive system to manage selfish nodes in vehicular networks. In this scheme, an incentive is given as a reward to the nodes for sharing different information. The provided information is related to the construction of roads, traffic congestion, and road accidents. In~\cite{33a}, the author proposed stable and reliable data dissemination that forward data to other nodes in the network intelligently. Sobin et al.~\cite{33b} proposed a selfishness and buffer-aware routing (SBR) to deal with the issue of selfishness. Yamini et al.~\cite{33c} proposed an advanced collaborative mechanism to deal with the problem of selfishness in Mobile Ad hoc network. Ganesan et al~\cite{33d} proposed Semi Markov process inspired selfish aware cooperative scheme (SMPISCS) for wireless sensor networks  In this scheme, selfish nodes are encouraged for cooperation in the network. Muhammed et al.~\cite{33e} presented Game Theory-Based Cooperation for Underwater Acoustic Sensor Networks. In this scheme, the selfish nodes are motivated for showing cooperation in the network. Terence et al.~\cite{33f} proposed behavior based routing misbehavior detection (BRMD) for wireless sensor networks to identify false advertiser node in the network. In this scheme, the sensor nodes are constantly monitored by the neighbor nodes. In smart and connected communities environment, safe and stable communication is required~\cite{33g}.  

Thus, SOS is a socially-oriented system that is used to forward data to the participating nodes in SCC depending on social interaction among nodes. Election scheme is used by SOS to encourage the selfish nodes for participation in a network. During the election, the elected heads motivate the selfish nodes for cooperation in a community. Nodes are given some incentive for their cooperative behaviors with other nodes that have greatly improved the performance of the network. Table~\ref{tab:nove} provide the comparison of SOS with some existing Schemes. 
\begin{table}[h]
\centering
\caption{Comparison of SOS with some Existing Schemes}
\begin{tabular}{p{3cm}p{4cm}p{4cm}p{4cm}}
\toprule
\textbf{Paper and Authors} & \textbf{Contributions} & \textbf{Limitations}& \textbf{Comparison with SOS}\\
\midrule
Umar et.al~\cite{27}& A game-theoretical reward-based system to manage selfish nodes in the network. Different types of cards are assigned to nodes according to their selfishness level in scheme.
&Individual importance factor problem: Only the individual importance of a node is considered for trust. 
Monitoring nodes can be selfish: This scheme only monitors the behaviors of the selfish node. & Collective Importance Factor (CIF) is computed to entitled any monitoring node to be cooperative or selfish.

It also monitors the behavior of the monitoring nodes in the network
\\ \\
Dias et al.~\cite{28}&
Proposed a cooperative watchdog system to tackle the issue of selfishness by assigning a reputation score to all nodes in the network.&Strict punishment issue: the node can not join the network once it has been punished.

It only detects Selfish nodes.& Node Selfishness is omitted in the proposed SOS scheme. Nodes are get warned for showing selfish behavior for the first time.
  \\ \\
Terence et al.~\cite{33f}& Proposed behavior-based routing misbehavior detection (BRMD) for wireless sensor networks to identify false advertiser node in the network.&  No punishment scheme is defined for selfish nodes by showing selfish behavior repeatedly.& A Proper punishment scheme is defined in the SOS scheme.
\\ \\
Lo et al.~\cite{34}& Proposed a multi-head clustering algorithm in vehicular ad hoc networks to handle the issue of selfishness& Weight tie Problem: when the weights of two nodes are the same, then there is no alternate criteria to elect heads in the election.& Cooperation is considered as the next criteria to elect heads during election  \\\\
\bottomrule
\end{tabular}
\label{tab:nove}
\end{table}
\section {SYSTEM MODEL}
\label{sys}

The proposed scheme SOS relies primarily on node involvement in the network. Nodes in the community have various performance-related activities, like forwarding of messages, monitoring, and tracking of nodes in the community. These activities are considered to be the main duties of the participating nodes in the election process. The nodes in the community are motivated to participate and function as a unit. One of the important features of the community-based node is to monitor the behavior of the neighbor node. This characteristic of a node provides complete control over the message transmitting and receiving. Therefore, an incentive system in the shape of reputation is suggested to encourage and stimulate the nodes to accomplish their tasks in the network by the proposed scheme. Nodes with malicious or selfish behaviors are motivated to engage in the voting system and act as cooperative nodes. Selfish nodes are also punished in the form of removal from the community for showing malicious behaviors repeatedly. This punishment message is broadcasted in the community regarding such nodes. Since the proposed system is divided into two phases such as election scheme and payment method. Thus, the nodes initiate involvement in the election process which establishes the election process. 

The community is controlled through the election process periodically. During the voting system, different heads are elected on the basis of higher weight, cooperation and a higher percentage of votes in the election. A node receiving a greater amount of votes, weight, and cooperation are elected as CH, second as MH, and third as IH. Election table is used to keep the records of all elected heads. The proposed scheme is applicable in IoT for smart and connected communities. Nodes are encouraged to cooperate with one another and due to this, selfishness is omitted which improves the livability, preservation, revitalization, and attainability of the community. The selfishness and cooperative nature of the nodes are determined by monitoring nodes through some trust value. The range of the trust value is in [0,1]. Nodes having a trust value greater than 0.5 is considered as cooperative and if the trust value is less than 0.5 then it is considered as selfish. Table~\ref{tab:not} lists the notations used in the proposed scheme. 
\begin{table}[h]
\centering
\caption{Notations}
\begin{tabular}{p{2cm}p{10cm}}
\toprule
Notation                 & Description and Explanation \\
\midrule

$x$ & It is the elected Community Head $CH$
during the election process on the basis of higher votes, weight, and cooperation
  \\

$cp_{x}$ & Contacts of the node $x$ with other nodes in the community
\\

$W_{x}$ & Weight of node $x$ shows the remaining resource it possesses  \\

$Vt_{x}$ & Node $k$ voted for community head $CH$ \\

$M_{x}$ & Monitoring node that constantly monitors the behavior of the nodes in the community
  \\

$P_{f}$ & Fixed Payment that is given to nodes for their cooperative behaviors
 \\

$F_{b}$ & Fixed Payment for each node that participates in the community election \\

$\Psi_{x}$ & Per member payment    \\

$IF_{x}$ & Importance Factor shows the honesty of node $x$  \\

\bottomrule
\end{tabular}
\label{tab:not}
\end{table}

\subsection{Election Participation Payment}

The proposed system SOS forward information in the community. The method starts with the involvement of nodes in the voting phase. The method of community formation and maintenance is an outreach for this article and therefore not discussed here. Table~\ref{tab:intro} lists some of the important variables that are used in the proposed scheme.
\begin{table}[h]
\centering
\caption{Introduction of some key variables}
\begin{tabular}{p{2cm}p{10cm}}
\toprule
Variables                 & Description \\
\midrule
$Weight$ & Weight is the number of remaining resources a node possesses. 
\\
$Cooperation$ & Nodes are regarded as cooperative if they make contacts with the more nodes in the community i.e $cp_{x} > k$ where $k=\lceil \frac{n}{3}\rceil$. 
\\
$Selfishness$ & To measure selfishness of the node, nodes with contact $cp_{x}\leq k$ are declared as selfish. \\

$Reputation$ &  Reputation of the monitoring nodes is calculated on its honest behavior in the network.\\

$CIF Rule$ & The Collective Importance Factor preludes the monitoring nodes from making prejudice decision about the node having a mutual relationship prior to it. \\

\bottomrule
\end{tabular}
\label{tab:intro}
\end{table}

\subsubsection{Election Procedure}

The first phase of SOS conducts the election process based on certain eligibility criteria. Two properties such as weight and cooperation of nodes are eligibility requirements. Nodes are nominated for the election that has greater (remaining) weight and cooperation. The weight is simply the total amount of resources a node possesses. 

Energy Ratio ($E_{x}$): The SCC devices are resource-constrained. Therefore, energy is also limited. Let $E_{x}$ be the energy and is provided by:
\begin{equation}
    E_{x}=\frac{Er_{x}}{Emax_{x}}\times 100\%
\end{equation}
Where the current remaining energy of node $x$ is $Er_{x}$ and highest energy of  node $x$ is $Emax_{x}$.

Buffer Ratio ($B_{x}$): The space in buffer is decreased by storing more data in it. Where $B_{x}$ is the proportion of the remaining buffer that reflects the node position in the buffer. The remaining status of the buffer is provided by:
\begin{equation}
B_{x}=\frac{Br_{x}}{Bmax_{x}}\times 100\%
\end{equation}
Where $Br_{x}$ is the remaining buffer of the node $x$ presently and $Bmax_{x}$is the highest buffer.

Remaining Time-To-Live $(TTL)$: $TTL$ is related to the delivery of a bundle. Each node should forward bundle before its $TTL$ is expired. Node is not considered for payment after its $TTL$ is expired. The status of node in the form of $TTL$ is given by:
\begin{equation}
T^{x}_{{ID}_{m}}=\frac {TTLr_{x}}{TTL}\times100\%
\end{equation}
where $T^{x}_{ID_m}$ is the remaining $TTL$ of bundle $ID_{m}$ carried by node $x$ presently and $TTLr_{x}$ is the remaining $TTL$ of the bundle. 

Node Degree $ND_{x}$: This indicates the number of nodes as neighboring nodes in the node $x$ transmission range.
\begin{equation}
ND_{x}=\sum_{y\in n,y\neq x} \big\{ y\mid dis(x,y)<T_{range}\big\}
\end{equation}

Relative Distance $(RD_{x})$: By~\cite{34}, the SCC nodes have certain features of how near they are to each other. Each node computes its own proximity to the mean distance and is provided by the formula given below:
\begin{equation}
RD_{x}=\mid _{pos} - \omega_{pos}\mid=\sqrt{((X_{x}-X_{\omega})^2+(Y_{x}-Y_{\omega})^2)}
\end{equation}
In the given equation, $x_{pos}$ indicates the location of $x$, the mean location of any node with its neighbors of $x$ is depicted by $\omega_{pos}$ , $(X_x,Y_x)$ is the $X$ and $Y$ coordinates of node $x$ and $(X_{\omega},Y_{\omega})$, shows the coordinate of $\omega$ position.
Once the results of all five characteristics are computed, their weight is determined by:
\begin{equation}
W_{x}=E_{x}.wt_{1}+B_{x}.wt_{2}+T^{x}_{{ID}_{m}}.wt_{3}+ND_{x}.wt_{4}+RD_{x}.wt_{5}
\end{equation}
where $W_{x}$ is the node $x$ weight, the weights $(wt_{1}, wt_{2}, wt_{3},wt_{4},wt_{5})$ are randomly chosen, where the absolute weight is equivalent to 1 comparable to ~\cite{35}. The scenario where the weight of two node is same, then cooperation is adopted as the next criteria for nomination of nodes for election. Cooperation is provided by:

\begin{align}
cp_{x}=\sum_{n\in N}Rc_{x}(n), if cp_{x}>k
\end{align}

Where $k=\lceil \frac{n}{3}\rceil $, $cp_{x}$ is the node $x$ cooperation, all nodes in the community is $n$ and $Rc_{x}$ is node $x$ total contacts. Nodes are regarded as cooperative if they make contacts with the more nodes in the community i.e $cp_{x} > k$. To measure selfishness of the node, nodes with contact $cp_{x}\leq k$ are declared as selfish. Node with greater weight and cooperation is nominated as an election candidate. The node makes communication in its range for a short duration. There is a chance that the information given by the node may be incorrect in terms of its weight and cooperation. A node may illustrate its weight as underweight and overweight. The underweight illustration will prevent it from being elected as community head and the overweight will offer it some opportunities to become a leader of the community. VCG model is used to develop and increase the trust behavior of the nodes within the community. The aim of this model is to expose the incorrect information regarding the node weight. Algorithm~\ref{alg1} provides the details of the election procedure. In this algorithm, nodes are nominated for election based on two characteristics weight and cooperation. The nodes having higher weight and cooperation are nominated for election.  After nomination, heads such as community Head $CH$, monitoring Head $MH$, and Incentive Head $IH$ are elected in the election by getting a higher number of votes. All the nodes that participated in the election are also awarded some sort of payment. Election table is used which contains all the record of the nodes that participated in the election. 

\begin{algorithm} 
\caption{Election procedure \& Payment to participating nodes} 
\label{alg1} 
\includegraphics[]{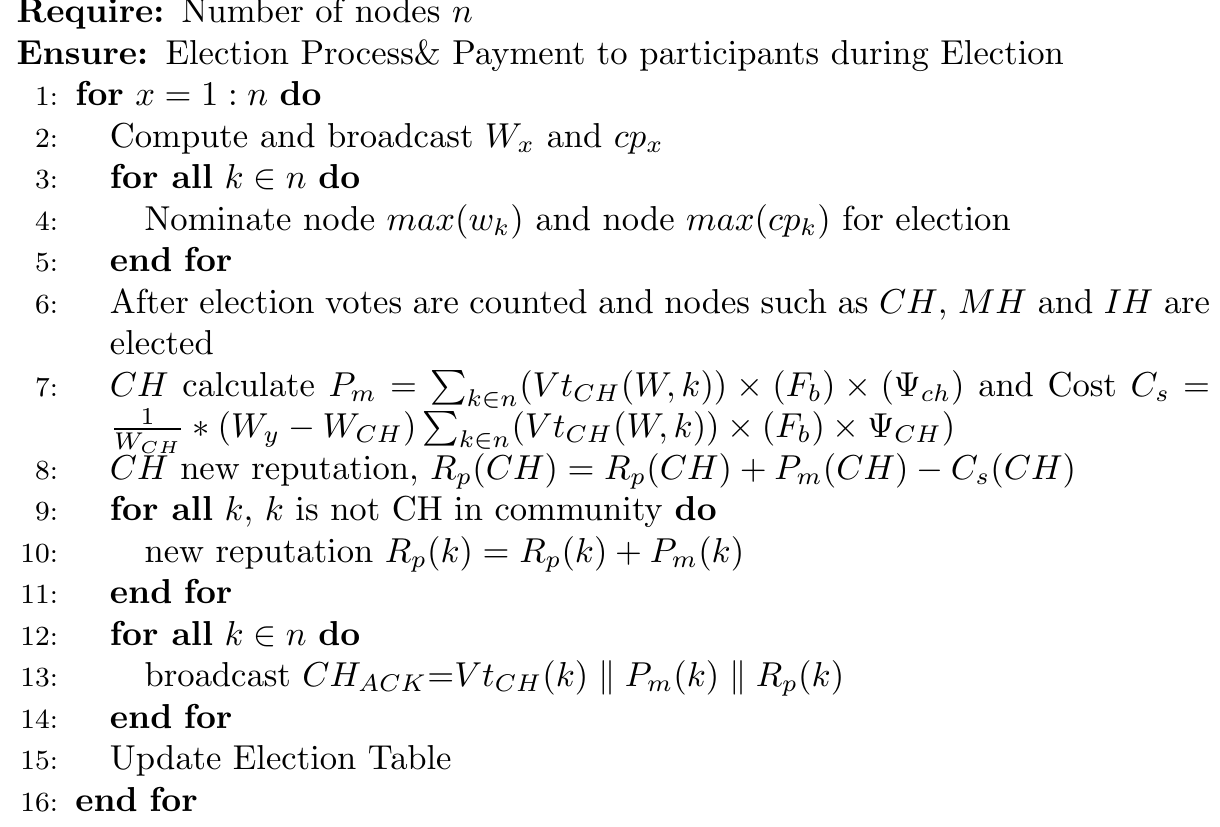}

\end{algorithm}


\subsubsection{Vickrey, Clarke, and Goves (VCG) Approach}

Vickrey, Clarke, and Groves (VCG) is a helpful technique that utilizes game theory tools. This model is used to demonstrate the behaviors of all nodes within the network and also encourage the node to tell the truth~\cite{36}. Let $P$ be all the members of VCG model, where each member $m\in\{{1,2,3,...n}\}$ has some personal information $\theta_{m}$, known as member type. Let $Z_{m}\in\Theta$ be any strategy that a member $m$ can use to input in a mechanism. Let $P=\{{P_{1},P_{2},P_{3},...,P_{n}}\}$ be the specific payment vector. To calculate explicit payment vector $P=\{{P_{1},P_{2},P_{3},...,P_{n}}\}$, the VCG model takes input from all the members to generate overall output $O=O\{{Z_{1},Z_{2}...,Z_{n}}\}$. The output generated shows the preference of each node as cost function $C_{m}=(\Theta,O)$. The handling of such information shows the usefulness of each member as calculated by cost function $U_{m}=P_{m}-C_{m}(\theta_{m},O)$.

To cope with the features of SCC, a slight modification is made to the present model in this article. Previously, only the energy level of a node was considered as its persona data. To illustrate the energy level of a node, the truth-telling behavior of the VCG model was used~\cite{37}. Here in the proposed scheme, the energy level of a node and as well as some other parameters such as buffer ratio, message $TTL$, relative distance and node degree are considered as the weight of a node. The individual personal information of a node is the weight of a node. Every node is awarded a real number called reputation value that depends on the reward or penalty a node receives from the community head. Nodes reputation improves or reduces after every voting process and is dependent on nodes cooperation within the network.

\subsubsection{Post Election Payment Based on VCG Model}

There are $n$ nodes in the game. Each node is a community player. The nodes in the contest or game need to disclose their weight to initiate the electoral process. In the election process, some nodes are elected as heads and recognize others as participants. Payment in the shape of reputation is made to both elected heads and participants. Every node in the contest tries to improve its reputation $R$. Higher reputation nodes get more network utilities. A reputation table known as $RTable$ is maintained by each node in the community. This table has all the details regarding the reputation of the neighbors and it is modified whenever necessary. Algorithm~\ref{alg2} provides the details of the payment process and operations of all heads elected in the election. In this algorithm, the monitoring nodes are assigned the responsibility to constantly check the behaviors of neighbor nodes in the community. The incentive head is responsible for making payment to the nodes. Nodes are awarded some incentive for showing cooperation in the community. But for showing selfish behavior in the community, nodes are also punished. Sometime Monitoring nodes can also be selfish. In such cases, $CH$ also computes the Importance factor of all monitoring nodes.

\begin{algorithm} 
\caption{Operational Phase \& Packet forwarding Payment Process} 
\label{alg2} 
\includegraphics[]{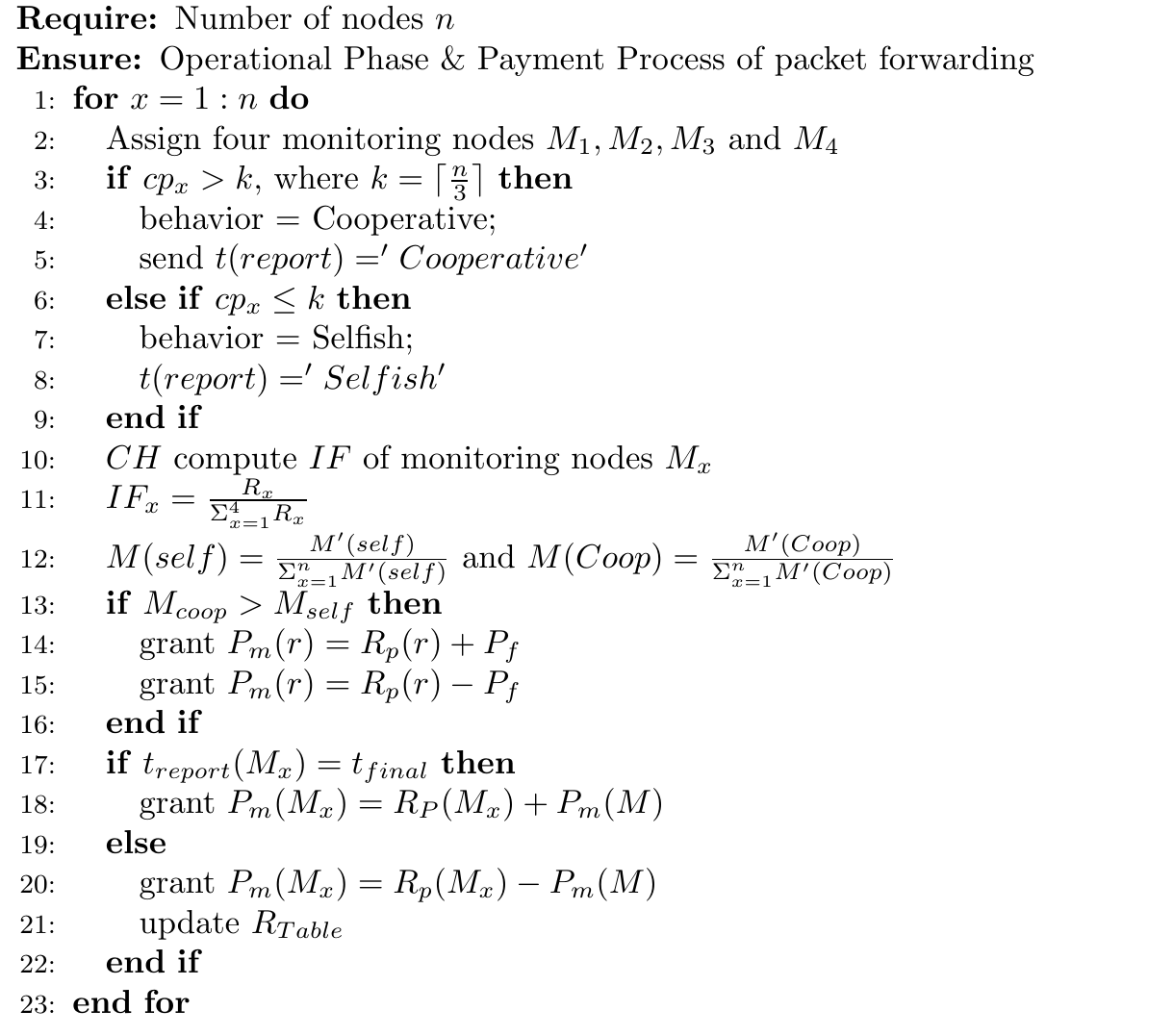}

\end{algorithm}


\paragraph{A. Community Head $(CH)$ Payment}

After the completion of the election process, every node in the network gets its payment. The payment to the $CH$ is made based on the votes given to it by the participating nodes. All the nodes in a community election process that voted for $CH$ are awarded incentive that is considered as the cost of the $CH$ elected in the election. The cost vector that is actually the weight of a node is expressed by $W_{1}, W_{2},....,W_{n}$, where $n$ is the number of nodes in total. The differentiation between receiving and making incentive is the $CH$ profit.
\begin{align}
P_{m}(x)=\sum_{k\in n}(Vt_{x}(W,k))\times(F_{b})\times(\Psi_{x})
\end{align}
Where $(Vt_{x}(W,k)$ in the election scheme generates particular value (equals 1 if $k$ votes for $x$, 0 is produced otherwise). The $IH$ also determines particular fixed budget$F_{b}$ for each node involved in election (this payment is well-know and fixed to all nodes) and $\Psi_{x}$ is node payment as provided below:
\begin{align}
\Psi_{x}&=W_{x}+\frac{1}{\sum_{k\in n}Vt_{x} (W,k)}
\times(\sum_{l\in n}(W_{y})).\notag\times{\sum_{k\in n}Vt_{y}}(W|W_{x}=\infty,k)-{\sum_{y\in n}}(W_{y})\\
{\sum_{k\in n}}(Vt_{y}(W,K))
\end{align}

\paragraph{B. Community Members Payment}
On the basis of fixed payment $F_{b}$, the absolute cost of the nodes is shared among all the nodes by the $CH$ (nodes that gave the vote). The cost function $Cost_{x}$ determined by $CH_{x}$ is given under:
\begin{align}
 C_{s}(x)=\frac{1}{W_{x}}\ast(W_{y}-W_{x})\sum_{k\in n}(Vt_{x}(W,k))\times(F_{b})\times\Psi_{x}
\end{align}
where $W_{x}$ and $W_{y}$ indicate the maximum and second maximum nodes weights of the node involved in the election process. The elected heads in a community deducted the absolute cost from their payment to calculate their own reputation.
\begin{equation}
  R_{p}(x) = P_{m}(x) - C_{s}(x)
	\label{eq01}
\end{equation}
The absolute cost of the nodes is shared among them depending on the nodes reputation. The $CH$ announces a payment to member nodes through some $CH_{ack}$ notification. The messages are signed and checked using standard message authentication. Each of the nodes updates the $RTable$.

\subsection{Packet Forwarding Payment in a Community}
The $CH$ and Gateway nodes are constrained for forwarding of messages only acting as relay nodes. The relay nodes may not forward some packets by showing selfish behavior. Such selfish behaviors of a node in SCC have undesirable effects on community performance. It also manipulates community nodes disconnection and added to the percentage of packet drop ratio. Each node receives an incentive for message forwarding in the form of reputation~\cite{35}. The monitoring nodes that control the function of relay nodes make the payment scheme more efficient.

\subsubsection{Payment for Relay Nodes in a Community}
Payment is made to the node for each packet forwarding in the proposed scheme. The incentive head makes a fixed payment $p_{f}$ to the nodes in the network. This $P_{f}$ is made on the cooperative behavior of the nodes. Therefore, the monitoring system is presented, that collects impervious of all monitoring nodes. These evidences contribute to the computation of the decision-making behavior of the relay nodes.

\paragraph{A. Collective Trust of Monitoring Nodes in a Community}

In the proposed scheme, a relay contains four monitoring nodes. In these four monitoring nodes, one of them is monitoring head that is elected in the election process. The other three monitoring nodes are participating nodes picked in a round-robin mode. A packet hash ~\cite{38} is generated by each node to maintain the packet genuine and prevent the packet from being changed by the forwarder. Furthermore, the hash score of the forwarded packet will be verified when the packet arrives at the next relay node. It ensures that the packet sent will be consistent whenever the hash results match. In a case, when hash results do not match, then the relay node is labeled as selfish and collect a negative reward. Each node keeps a record of the forwarded packets in its buffer.
These packets will be sent next after some expected lifetime. The monitoring node generates a trust report regarding the behavior of nodes after certain threshold time period. The trust report specifies nodes intentions to forward messages. The trust report shows the behavior of the nodes as selfish or cooperative. The four monitoring nodes send a report to $CH$ to compute the trust value. If the Collective trust assessment of the cooperative nodes surpasses the malicious conduct, the forwarder is labeled genuine and receives favorable payment (reputation) otherwise selfish nodes will receive negative payment after repeatedly showing selfish behavior (punishment). To obtain comparable outcomes (avoid inconsistent results), the Collective Importance Factor (CIF) principle is presented to compute the trust depends on evidence from distinct nodes.

\paragraph{B. Use of Extended Dempster-Shafer to Merge Evidences }

The Dempster-Shafer uses mathematical formulas to resolve uncertainty situations in a network~\cite{39},~\cite{42}. This model is primarily used in finding routing attacks in mobile ad hoc networks~\cite{36} and for calculation of collective trust. The evidence theory is calculated by $\delta$, which is a frame of judgment and probability assignment function BPA~\cite{35}. A frame of judgment $\delta$ shows a mutually exclusive and exhaustive hypothesis, showing only one of them is true. BPA function shows $b:2^\delta\longleftarrow[0,1]$ satisfying two of the condition:
\begin{equation}
b(\emptyset)=0
\end{equation}
and
\begin{equation}
\sum_{A\subseteq\vartheta} b(A)=1
\end{equation}
Where $\vartheta$ is null set having A is any subset of $\delta$ to calculate two BPA functions $b_{1}$ and $b_{2}$, the DS theory gives the following rule:
\begin{equation}
b(C)=\frac{\sum_{A\cap B=C}b_{1}(A)b_{2}(B)}{1-\sum_{A\cap B=\emptyset}b_{1}(A)b_{2}(B)}
\end{equation}

The limitations of DS theory has that it treat all the evidences equal and the priorities of the evidences are not considered. ~\cite{40} has introduced the importance factor IF in the proposed Extended Dempster-Shafer (EDS) rule. Where IF is a real number calculated on the basis of the importance of evidence. ~\cite{40} defines basic probability assignment rules for two importance factor $IF_{1}$ and $IF_{2}$ having $EV_{1}$ and $EV_{2}$ evidences:
\begin{equation}
b(C,IF_{x}, IF_{w})=\frac{\sum_{A_{x}\cap B_{x}=C}\bigg[( b_{1}(A_{x})^\frac{IF_{x}}{IF_{w}}b_{2}(B_{w})^\frac {IF_{w}}{IF_{x}}\bigg]}{\sum_{C\subseteq\theta,C\neg=\emptyset},\sum_{A_{x}\cap B_{x}=C}\bigg[( b_{1}(A_{x})^\frac{IF_{x}}{IF_{w}}b_{2}(B_{w})^\frac {IF_{w}}{IF_{x}}\bigg]}
\end{equation}

However, in some situations, both DS and EDS theories generate output which is irrational~\cite{41}.

\paragraph{ C. Trust Calculation Based on Collective Importance Factor}

The Collective importance factor preludes the monitoring nodes from making prejudice decision about the node having a mutual relationship prior to it. It may declare the node as selfish and punish it. Thus, an importance factor that distinguish between honest and dishonest monitoring nodes in a community is essential. The reputation of the monitoring nodes is calculated on its honest behavior in the network. Monitoring nodes importance factor is its honesty. The $IF$ of monitoring nodes $x$ is equal to the reputational score over the actual reputational score of all monitoring nodes in the community. $IF_{x}$ indicates the importance factor of monitoring node x and $R_{1},R_{2},R_{3}$, and $R_{4}$ are the four monitoring nodes taking part in a relay. 
\begin{align}
IF_{x}=\frac {R_{x}}{\Sigma^{4}_{x=1} R_{x}}                       
\end{align}
Implicitly if any node $x$ report the behavior of any node $y$, the accurate or genuine judgment is equal to the importance factor $(IF)$ of a node reporting the behavior of a node. Any node $x$ having an $IF_{x}$ report that node $x$ is cooperative.
\begin{align}
M_{x}(Cooperative)=IF_{x}
\end{align}
\begin{align}
M_{x}(Selfish)=1-IF_{x}
\end{align}
In a similar manner, if any node $w$ reported $k$ as non-cooperated or selfish then
\begin{align}
M_{w}(Selfish)=IF_{w}
\end{align}
\begin{align}
M_{w}(Cooperative)=1-IF_{w}
\end{align}
	
The CIF rule calculates the collective trust as under.
Suppose $(A,b_{1}), (B,b_{2})....(N,b_{n})$ are two discrete evidence produced by $n$ watchdog having $IF_{1},IF_{2}$ and $IF_{n}$ as an importance factor. $EV_{1}, EV_{2}....EV_{k}$ shows $k$ combination of elements in $\delta$ given in Eq(21). For each $EV_{w}$, $w \in k$ we associate a value $b'(EV_{w})$ as, 

\begin{align}
b^{\prime}(EV_{w})=\displaystyle\sum_{x=1}^{n}b_{x}(EV_{w})^\frac{IF_{x}}{\Sigma^{n}_{i=1, i\neq x}IF_{i}}-\prod_{x=1}^{n}b_{x}(EV_{w})^\frac{IF_{x}}{\Sigma^{n}_{i=1, i\neq x}IF_{i}}
\end{align}

Finally, BPA is assigned to $EV_{x}$ as,

\begin{align}
b(EV_{w})=\frac{ b'(EV_{w})}{\Sigma^{n}_{x=1}b'(EV_{w})}
\end{align}

\paragraph{D. Descriptive instance of Collective Importance Factor Rule}

Consider four monitoring nodes $M_{1},M_{2},M_{2},$ and $M_{4}$ with a reputation score of 70, 30, 10 and 40 respectively produce a forwarder trust report on $node_{1}$. The $node_{1}$ is cooperative according to the report of $M_{1}$. However, $M_{2}$, $M_{3}$ and $M_{4}$ report $node_{1}$ is not cooperative. So, the importance factors $IF_{1}$, $IF_{2}$, $IF_{3}$, and $IF_{4}$ of the four monitoring nodes are calculated for their total reputation score
 as, $\frac{70}{100}=0.7$,  $\frac{30}{100}=0.3$, $\frac{10}{100}=0.1$, and $\frac{40}{100}=0.4$ respectively. Thus,\\
 
$M_{1}(cooperation)=0.7$    \space\space\space\space\space $M_{1}(Selfishness)=0.3$ \\     

$M_{2}(Selfishness)=0.3$  \space\space\space      $M_{2}(Cooperation)=0.7$ \\

$M_{3}(Selfishness)=0.1$  \space\space\space      $M_{3}(cooperation)=0.9$ \\

$M_{4}(Selfishness)=0.4$     \space\space\space      
$M_{4}(cooperation)=0.6$ \\

Subsequently, the collective trust value on node 1 is computed as, \\

$M'(Selfishness)=(0.3)^\frac{0.7}{0.3}+(0.3)^\frac{0.3}{0.7}+(0.1)^\frac{0.1}{0.9}+(0.4)^\frac{0.4}{0.6}-(0.3)^\frac{0.7}{0.3}(0.3)^\frac{0.3}{0.7}(0.1)^\frac{0.1}{0.9}(0.4)^\frac{0.4}{0.6}=1.95$\\

$M'(Cooperation)=(0.7)^\frac{0.7}{0.3}+(0.7)^\frac{0.3}{0.7}+(0.9)^\frac{0.1}{0.9}+(0.6)^\frac{0.4}{0.6}-(0.7)^\frac{0.7}{0.3}(0.7)^\frac{0.3}{0.7}(0.9)^\frac{0.1}{0.9}(0.6)^\frac{0.4}{0.6}=2.73$\\

Therefore, $M(Selfishness)$ and $M(Cooperation)$ can be computed as,\\

$M(Selfishness)=\frac{1.95}{1.95+2.73}=.416$

$M(Cooperation)=\frac{2.73}{1.95+2.73}=.583$ \\ 

The evidences provided by all the nodes in the network will decide the honesty of the forwarder for making payment. CIF principle decides the selfish and cooperative nature of nodes. The CIF rule states as if three monitoring nodes declare a node selfish but the final trust calculated is less than the fourth one, the node will still be cooperative. This implies the node having cooperative nature for a long time has its importance. As of final calculation, the new reputation is computed from the nodes given incentive and contemporary reputation, and it is broadcasted by the $CH$.

\subsubsection{Monitoring Nodes Payment in a Community}

Monitoring nodes are paid for submitting trustworthy reports. The $CH$ is making the payment to the monitoring node based on some trust score. Monitoring nodes trustworthiness is determined by the final score. For $(P_{m}(M)>0)$, implies that the monitoring nodes are trustworthy as the final trust score matches the trust report. For $(P_{m}(M)<0)$, implies that the final trust value has a deviation from the trust value and termed the node as misbehavior monitoring node.

Some changes have been made to the new model to handle the devices in the community. Therefore, the nodes weight is regarded as individual personal information that is one of the eligibility requirements for participation in the process of election. In addition, a nodes reputation is  a real number allocated to each
participating node calculated on node behavior. This value varies according to the truth-telling behavior of the nodes in the network.

\subsubsection{Reputation Carry and Forward}

At first, all nodes have zero reputation as they join the network. The reputation of the nodes varies during the electoral phase. This is reported by the $CH$ periodically to the nodes in the network. Node switches its position to $GW$ as it gets updates from $CH$. The node constantly updates the $RTable$ with other $CH$ as it gets updated in $Rtable$ in either community. Community head in the network confirms the accuracy and integrity of the information transmitted by $GW$ node. It affirms that the $CH$ knows about the information about two hope communities in both directions. For instance, there are three communities namely $A,B,$ and $C$. Suppose a new node $X$ is joining the community $B$. Community $B$ apprehend all the reputation score of all the nodes in $A$ and $C$ as per the proposed scheme. Thus, the reputation score of the node $X$ is recognized to the community $B$ even before it joins $B$. If the $CH$ is unsure of the new node reputation score, a new node will be recognized as a new node within a network.

\subsection{Selfish Nodes Punishment in a Community}

Community head stimulates the selfish nodes to take an active part in the election process. This participation in election process labels them as cooperative nodes. The head of the community can punish the nodes with selfish behavior in three ways. The community head stimulates the node to cooperate after being selfish for the first time but in such case, no incentive is awarded. Secondly, a node receives negative payment after warning from the community heads. Finally, the community heads can remove the selfish nodes from the community as punishment for a certain time. But sometimes a node can enter the network again and act cooperatively, so in such scenario, the negative payment should be paid first.
\section {PERFORMANCE EVALUATION}
\label{per}

This section outlines the performance of SOS in Network Simulator NS-2.34~\cite{43},~\cite{44} by comparing it with current algorithms. NS-2 is a network simulator for both wired and wireless networks. MANET routing protocols can be implemented in NS-2. Here, the simulation setup and metrics are presented first and then simulation results are discussed.

\subsection{Simulation Setup}

The simulation is performed in six modules. The first module shows how the reputation of nodes changes with selfish nodes variation. It is presented in the simulation during the election process. The second module show nodes behavior changes with respect to its variation in reputation. In the third module, the results for the CAIS protocol is obtained based on our setup. In the fourth module, we examined the SSAR protocol. The fifth module is about the proposed scheme. In CAIS, SSAR, and SOS, 4\% selfish nodes are injected in it. The results are compared and evaluated in the sixth module. The variations in the selfish nodes are used as tools to test all three protocols. The system is assumed to be normal under 0\% selfish nodes, which means that all the nodes are cooperative in nature. The selfish nodes are variable ranging from 4\% to 90\% in the network during the simulation. SSAR protocol is used as a benchmark. The simulation parameters are given in Table~\ref{tab:para}.

\begin{table}[h]
\centering
\caption{Parameters values for Simulation}
\begin{tabular}{@{}ll@{}}
\toprule
Parameter &         Values \\
 \midrule
 Area  &  $500\times500 m^2 $  \\
     Base Protocols	& CAIS, SSAR    \\
     Number of nodes&50  \\
     Node Distribution &Uniform\\
          Initial Energy	&90 \\
     $R_{x}$ Power & 0.3\\
   $T_{x}$ Power	&0.6\\
      Movement Trace	&OFF \\
      Malicious Activity &4\%,10\%, 25\%, 50\%, 75\%, and 90\%\\
    Comparison & CAIS, SSAR with selfish nodes 
		(Proposed Work)\\
                       Size of Packet Header&4 bytes\\
     Address Size	&4 bytes\\
     Max. number of messages/packet	&4bytes\\
                   Traffic Source&CBR\\
     Packet Protocol&TCP\\
\bottomrule
\end{tabular}
\label{tab:para}
\end{table}

\subsection{Metrics}

The simulation uses throughput, average delivery delay, average energy consumption, and packet delivery ratio (PDR) as metrics of performance in a network. Throughput~\cite{44a} is the number of successfully delivered packets to the total packets. The packet delivery ratio~\cite{32} is the successful delivery of the messages over generated messages. The average delivery delay~\cite{20} is defined as the time is taken by the message to reach its destination. Energy consumption~\cite{44b} of the individual nodes to energy consumption by the entire number of nodes is called average energy. SOS is compared with the following protocols: CAIS: A Copy Adjustable Incentive Scheme~\cite{20} and Social selfishness Aware Routing (SSAR)~\cite{32}. Both CAIS and SSAR are an incentive-based and social scheme that handles the issue of selfishness. Thus, these two schemes are used as a benchmark for SOS.

\subsection{Results and Discussion}

The nodes get payments through VCG model. Nodes involved in the election are paid for exhibiting cooperation in the network and become community head, monitoring head and Incentive head via the election process. The nodes with selfish behavior or action and not exhibiting the desired duty receive negative payments as punishment.

\subsubsection{Variation of Reputation}

\begin{figure}[ht]
\begin{subfigure}{.5\textwidth}
  \centering
  \includegraphics[scale = 1.1]{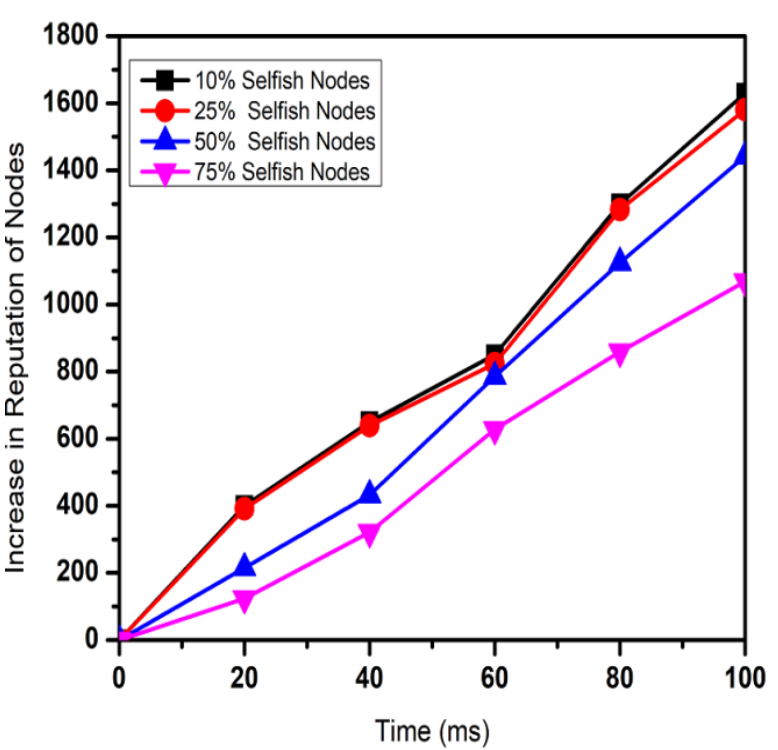}  
  \caption{Total reputation of nodes with modifying selfish nodes}
  \label{fig:subrep}
\end{subfigure}
\begin{subfigure}{.5\textwidth}
  \centering
  \includegraphics[scale = 1.08]{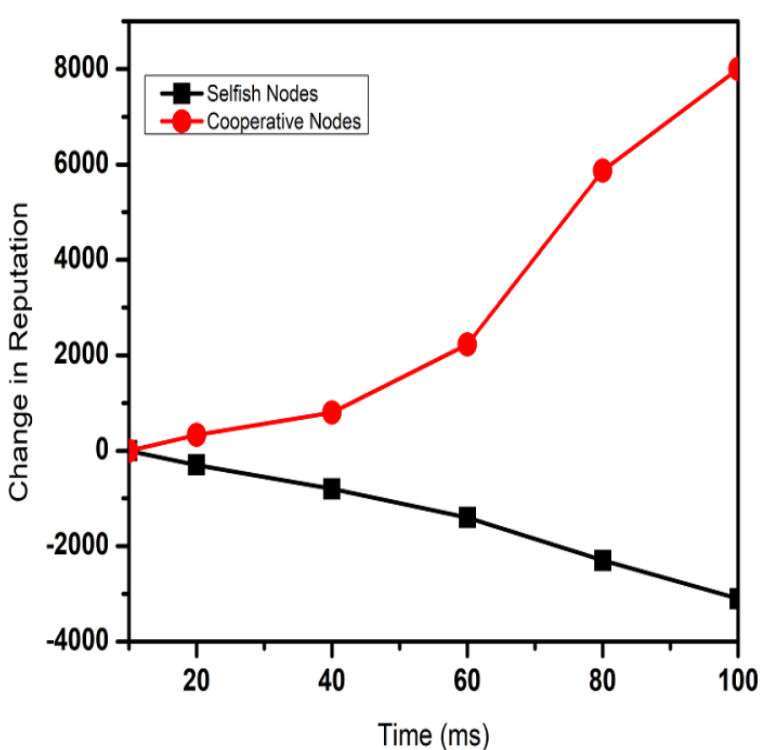}  
  \caption{Variation in Reputation over the simulation time}
  \label{fig:subchn}
\end{subfigure}
\caption{Total reputation of nodes with modifying selfish nodes \& Variation in Reputation over the simulation time}
\label{fig:fig1}
\end{figure}

The behavior of nodes has an effect on the reputation of the nodes. Reputation of nodes varies with its behavior. A node can be selfish and cooperative, determined by its behavior in the network. The variation in node behavior is shown in figure~\ref{fig:subrep} in the election process.

The final simulation results show that participated nodes are decreased as the selfish nodes increases during the election process. It also generates lesser payments to the selfish nodes by the heads of the community. Figure~\ref{fig:subchn} shows the behavior of the node in the election process. 
The reputation of the nodes in the network increases with an increase in cooperation and gets decreased as the number of selfish nodes increased. The ratio of reputation is higher in the cooperative nodes than the selfish nodes reputation as shown in the simulation results. Variation in the reputation ratio is due to the payment of incentives to the nodes that varies during the election process. It depends on the actively participating nodes in the network. Fixed payment to the relay node and monitoring nodes is made based on node behavior to forward messages to its neighbor. It implies that the nodes reputation increases with an increase in the participating nodes in the electoral process and number of the message forwarded by the node to its neighbor. The reputation of the nodes decreased when negative payments are made to the node. 
\subsubsection{Comparison for Injecting 4\% Selfish Nodes}

\begin{figure}[ht]
\begin{subfigure}{.5\textwidth}
  \centering
  \includegraphics[scale = 1.1]{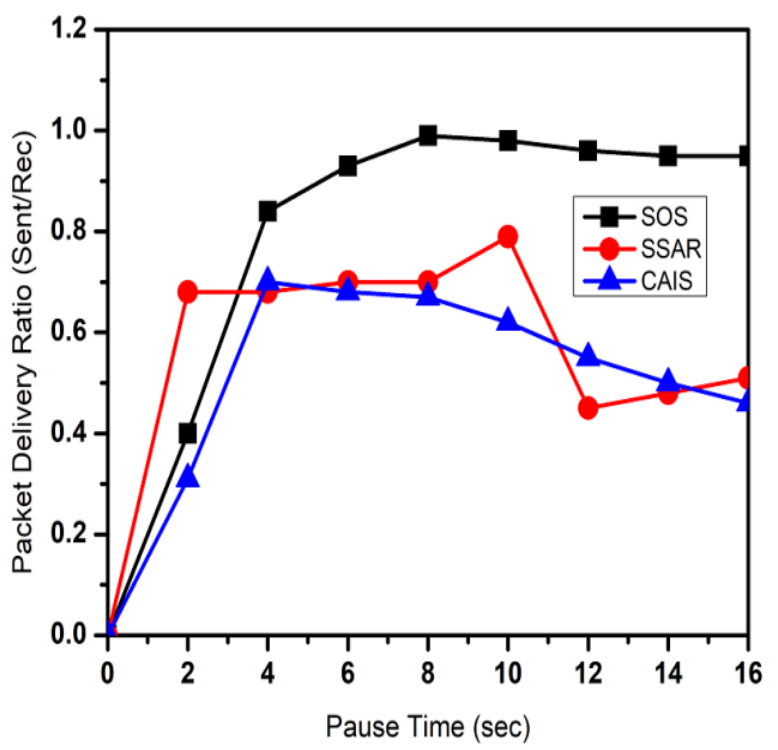}  
  \caption{Packet Delivery Ratio}
  \label{fig:pkt4}
\end{subfigure}
\begin{subfigure}{.5\textwidth}
  \centering
  \includegraphics[scale = 1.1]{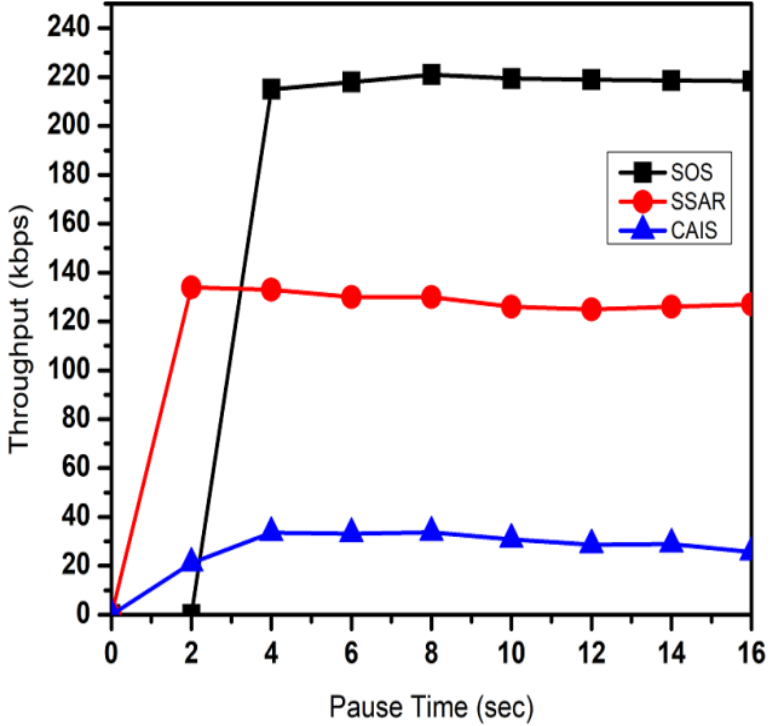}  
  \caption{Throughput}
  \label{fig:thr4} 
\end{subfigure}
\begin{subfigure}{.5\textwidth}
  \centering
  \includegraphics[scale = 1.1]{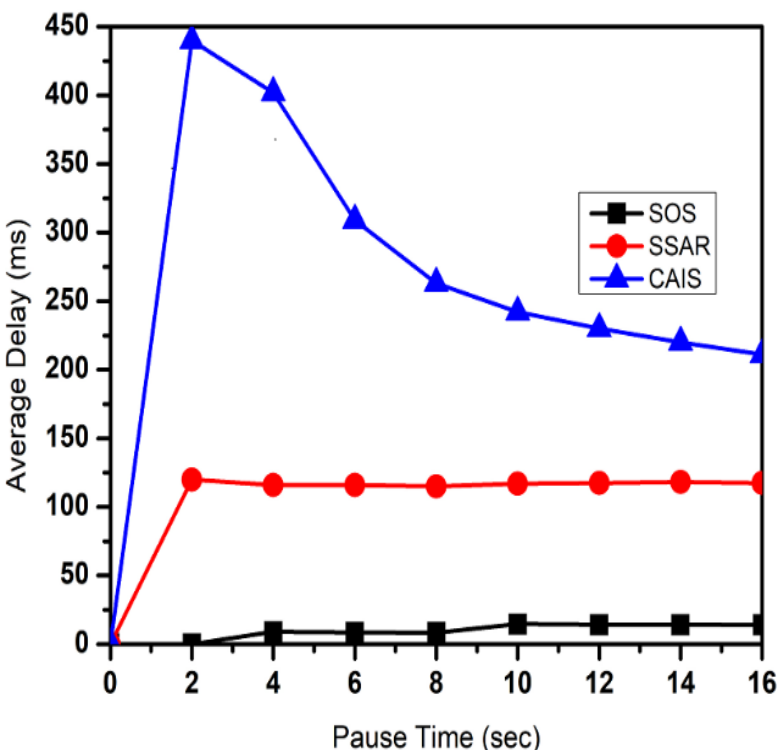}  
  \caption{Average Delivery Delay}
  \label{fig:del4}
\end{subfigure}
\begin{subfigure}{.5\textwidth}
  \centering
  \includegraphics[scale = 1.1]{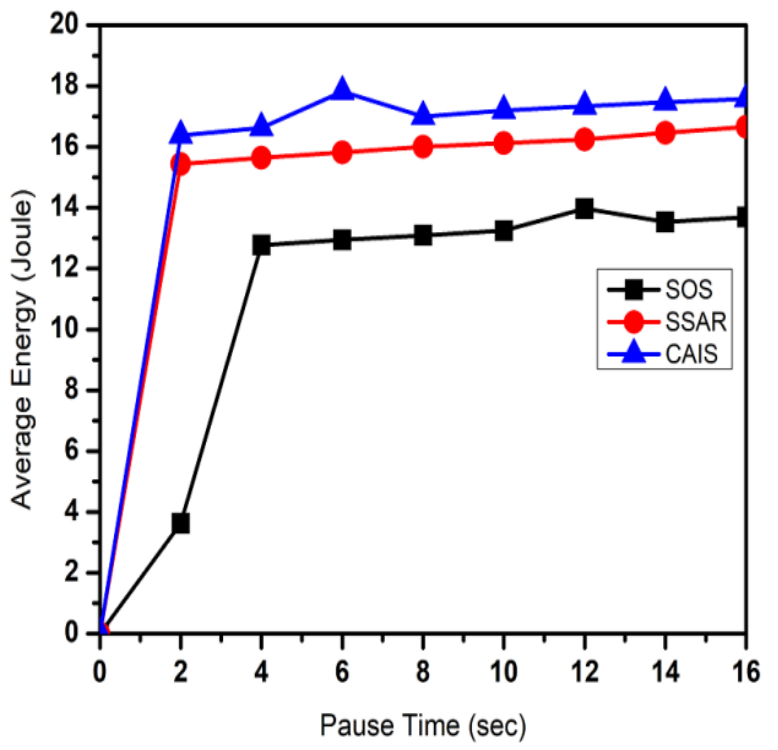}  
  \caption{Average Energy}
  \label{fig:ene4}
\end{subfigure}
\caption{Performance comparisons of the algorithms when 4\% of nodes are selfish}
\label{fig:fig2}
\end{figure}

The routing performance is studied in terms of packet delivery ratio, throughput, average delay, and average energy consumption when 4\% of nodes are selfish. Figure~\ref{fig:pkt4} shows the results of the performance metrics packet delivery ratio. The SOS technique gains the highest packet delivery ratios of the packets. At pause time 8 sec, the SOS has the highest packet delivery ratio of 0.9 (packet sent/rec) that is approximately 25\% and 27\% higher than SSAR and CAIS respectively. Figure~\ref{fig:thr4},~\ref{fig:del4} and ~\ref{fig:ene4} show the results of the performance metrics, throughput, average delivery delay, and average energy consumption. 
At pause time 8 sec, the throughput of SOS is 221 kbps that is 37\% and 78\% higher than SSAR and CAIS respectively. At pause time 10 sec, the average delay of SOS is 14.67 ms that is 50\% lower than CAIS and 23\% lower than SSAR. In addition, the average energy consumed by SOS is 12.77 joule at pause time 4 sec that is 14\% and 19\% higher than SSAR and CAIS respectively. The energy of the node is sometimes saved by not giving a due response to other nodes or by using a blacklisting mechanism. The SOS technique has comparatively high throughput, minimum delay, and energy consumption. It is due to the core reason that the SOS technique stimulates the selfish nodes in the network to participate in the network and effectively forward the messages. Comparing the results with SSAR, here the messages are forwarded on the node contact history and willingness level. The SOS technique takes some initial time (configuration and loading time) for the arrangements of all the factors involved in the simulation environment. The values of the simulation are not accurate for the initial two seconds. The values (Throughput and PDR) become consistent after pause time 2 sec. It can be shown in figure~\ref{fig:fig2} that the performance of SOS is better than the two existing techniques. 

The comparison of SOS, SSAR, and CAIS for 4\% selfish nodes for all performance metrics are shown in Table~\ref{tab:comp4}. It can be seen in~Table\ref{tab:comp4}, the SOS technique gains the highest packet delivery ratios of the packets. At pause time 8 sec, the SOS has the highest packet delivery ratio of 0.9 (packet sent/rec) that is approximately 25\% and 27\% higher than SSAR and CAIS respectively. At pause time 8 sec, the throughput of SOS is 221 kbps that is 37\% and 78\% higher than SSAR and CAIS respectively. At pause time 10 sec, the average delay of SOS is 14.67 ms that is 50\% lower than CAIS and 23\% lower than SSAR. In addition, the average energy consumed by SOS is 12.77 joule at pause time 4 sec that is 14\% and 19\% higher than SSAR and CAIS respectively~\ref{tab:comp4}.
\begin{table}[ht]
\begin{center}
\caption{Performance comparisons of SOS, SSAR, and CAIS for 4\% selfish nodes}
\label{tab:comp4}
\newcolumntype{b}{X}
\newcolumntype{c}{>{\hsize=.1\hsize}X}
\newcolumntype{d}{>{\hsize=.1\hsize}X}
\setlength{\extrarowheight}{2pt}%

\begin{tabularx}{\textwidth}{ c  cccc  d  dd  c  c  d  dd }
\toprule
&&\textbf{PDR}&&& \textbf{Throughput} &&&	\textbf{Avg.Delay}  &&&	\textbf{Avg.Energy}\\
\toprule
\textbf{Pause time}&\textbf{SOS} & \textbf{SSAR } &	\textbf{CAIS}  &	\textbf{SOS} &	\textbf{SSAR} & \textbf{ CAIS} &	\textbf{SOS} &	\textbf{SSAR } &	\textbf{CAIS} &	\textbf{SOS} &	\textbf{SSAR} &	\textbf{CAIS}\\
\midrule

0 &	0.0 &	0.0 &	0.0	& 0.0 &	0.0 &0.0 &0.0& 0.0&0.0&	0.0	& 0.0 & 0.0\\
2 &.4	&.68 &	.31& 0.0 & 134&21.08&0.0&120 &440&	3.6 &15.4&	16.3 \\
4 & .84 &.68&0.7&215& 133&33.6& 9.0 & 116&402&12.7 &	15.6 &16.6\\
6	&.93&0.7&.68&218&130 &33.2&	8.5&	116&309&	12.9&	15.8&	17.8\\
8 &	.99 &0.7 &.67& 221 &130&33.7&8.4& 115&263 &	13.0	& 16.0 & 17.0 \\
10 &.98 &0.7 &	.62	& 219.4 &126&30.9 &	14.67& 117&242&13.25& 16.12 & 17.2 \\
12 &.96 &.45 &	.55	& 218.9 &125&28.61&	14.16& 117.4&230 &	13.97& 16.24 & 17.3 \\
14 &.95&.48&.55	&  218.6&126&28.87&	14.1& 118.1&220 &13.53& 16.46 & 17.4 \\
16 &.95&.51&.46	& 218.4 &127&25.61&	14.05& 117.4&211 &	13.69	& 16.66 & 17.5 \\
\bottomrule

\end{tabularx}
\end{center}
\end{table}
\subsubsection{Influence of Different Percentage of Selfish Nodes}

The performance of SOS is compared with SSAR and CAIS for obtained network properties. The selfish nodes of 10\%, 25\%, 50\%, 75\%, and 90\% are injected and results are checked for packet delivery ratio, average deliver delay, throughput and average energy.

\begin{figure}[ht]
\begin{subfigure}{.5\textwidth}
  \centering
  \includegraphics[scale = 1.1]{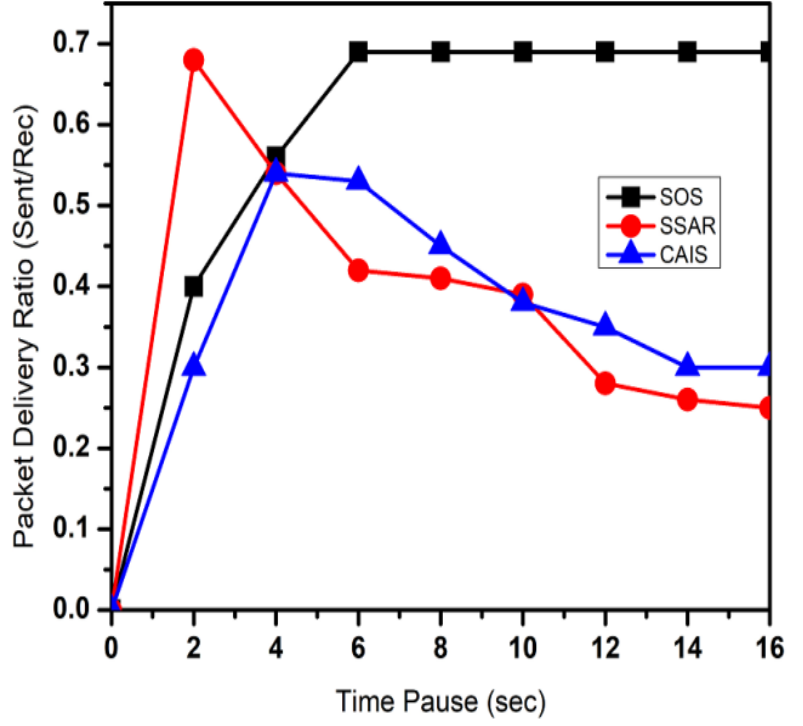}  
  \caption{Packet Delivery Ratio}
  \label{fig:pkt10}
\end{subfigure}
\begin{subfigure}{.5\textwidth}
  \centering
  \includegraphics[scale = 1.1]{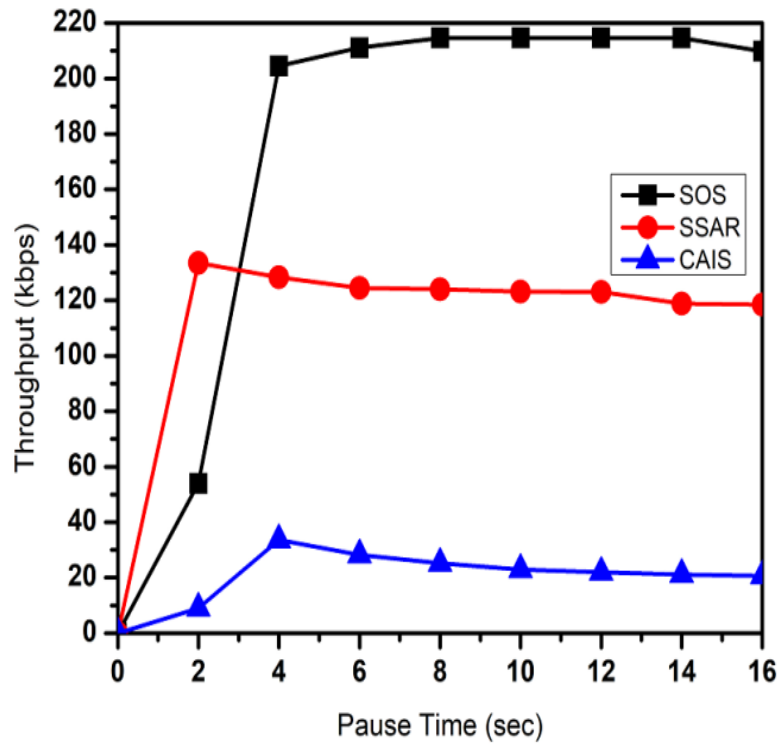}  
  \caption{Throughput}
  \label{fig:thr10} 
\end{subfigure}
\begin{subfigure}{.5\textwidth}
  \centering
  \includegraphics[scale = 1.1]{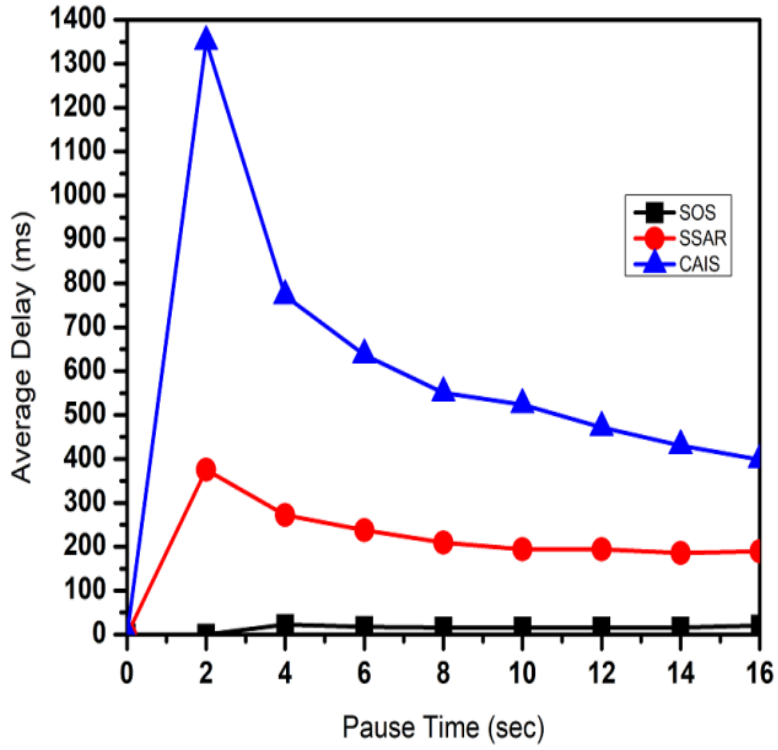}  
  \caption{Average Delivery Delay}
  \label{fig:del10}
\end{subfigure}
\begin{subfigure}{.5\textwidth}
  \centering
  \includegraphics[scale = 1.1]{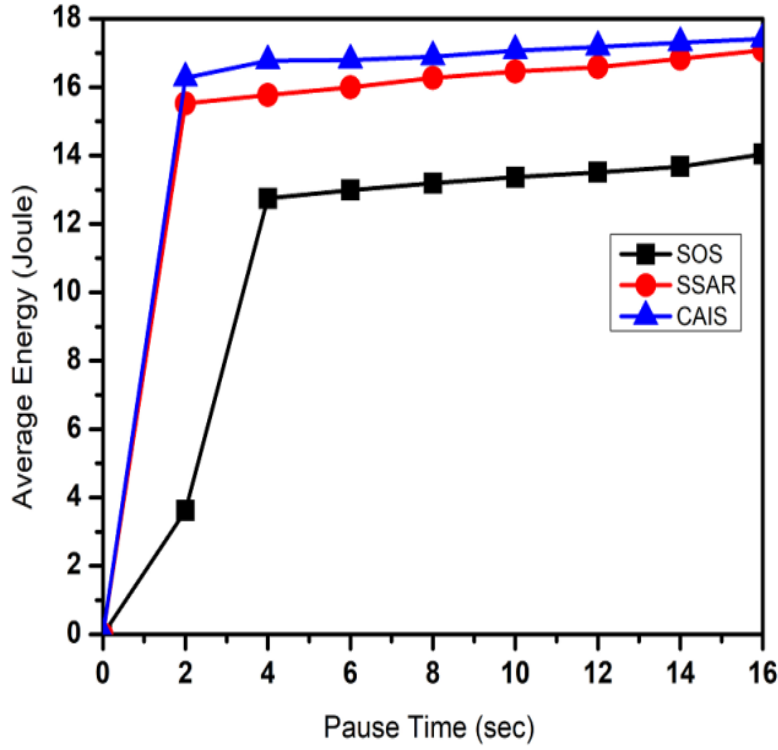}  
  \caption{Average Energy}
  \label{fig:ene10}
\end{subfigure}
\caption{Performance comparisons of the algorithms when 10\% of nodes are selfish}
\label{fig:fig3}
\end{figure}

By injecting 10\% selfish nodes in the network, SOS again outperforms CAIS and SSAR in terms of throughput, packet delivery, average energy and average delivery delay as shown in Figure~\ref{fig:fig3}.
At pause time 8 sec, the packet delivery ratio of SOS is 0.67 (packet sent/rec) that is almost 37\% and 31\% higher than SSAR and CAIS respectively as shown in Figure~\ref{fig:pkt10}. The throughput of SOS is 204.5 kbps at pause time 4 sec that is 34\% and 77\% higher than SSAR and CAIS respectively as shown in Figure~\ref{fig:thr10}. In addition, the average delay of SOS is 15.94 ms at pause time 8 sec that is 14\% and 38\% lower than SSAR and CAIS respectively as shown in Figure~\ref{fig:del10}. Similarly, the average energy consumed by SOS is 3.6 joule at pause time 4 sec that is 17\% and 23\% lower than SSAR and CAIS respectively as shown in Figure~\ref{fig:ene10}. 

\begin{figure}[ht]
\begin{subfigure}{.5\textwidth}
  \centering
  \includegraphics[scale = 1.1]{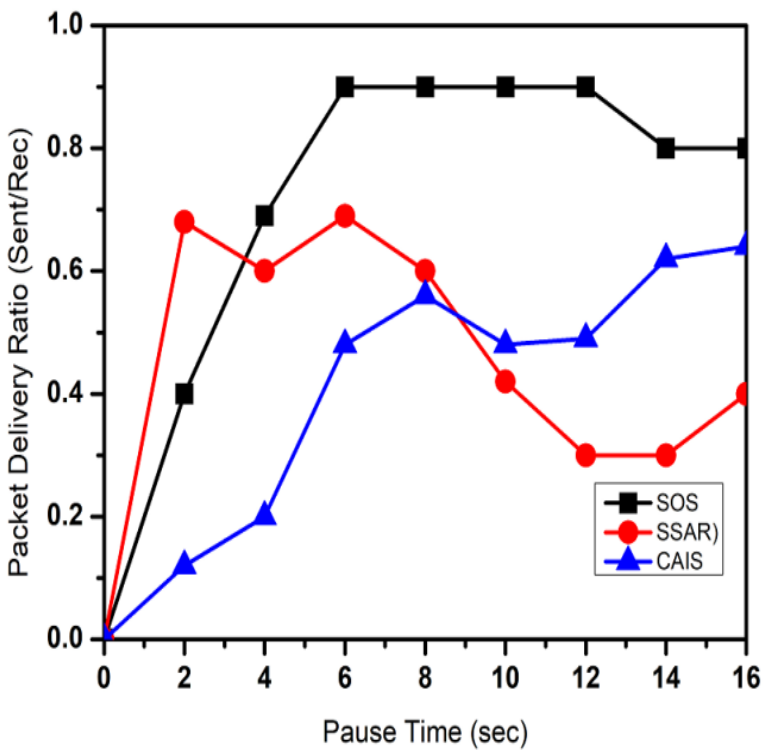}  
  \caption{Packet Delivery Ratio}
  \label{fig:pkt25}
\end{subfigure}
\begin{subfigure}{.5\textwidth}
  \centering
  \includegraphics[scale = 1.1]{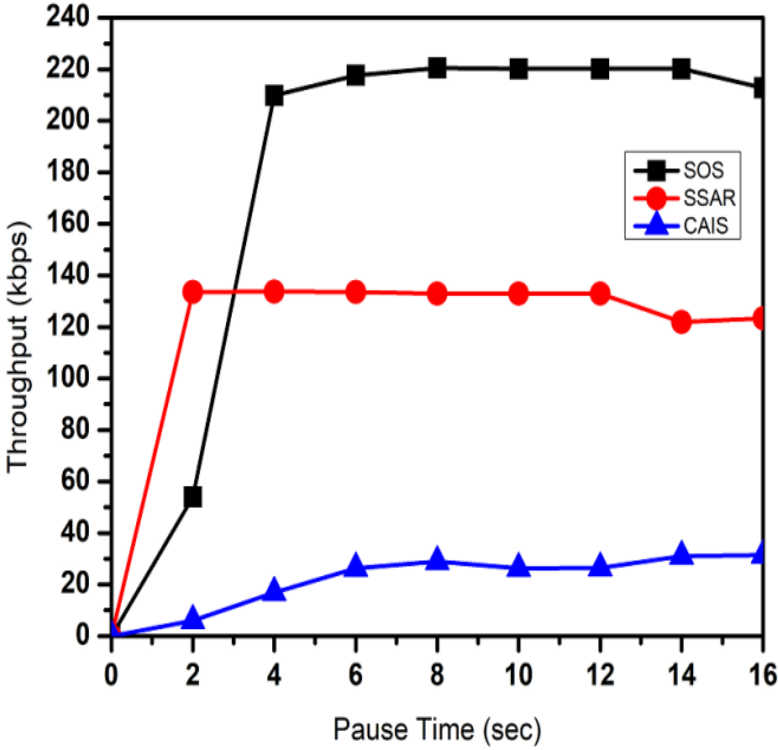}  
  \caption{Throughput}
  \label{fig:thr25} 
\end{subfigure}
\begin{subfigure}{.5\textwidth}
  \centering
  \includegraphics[scale = 1.1]{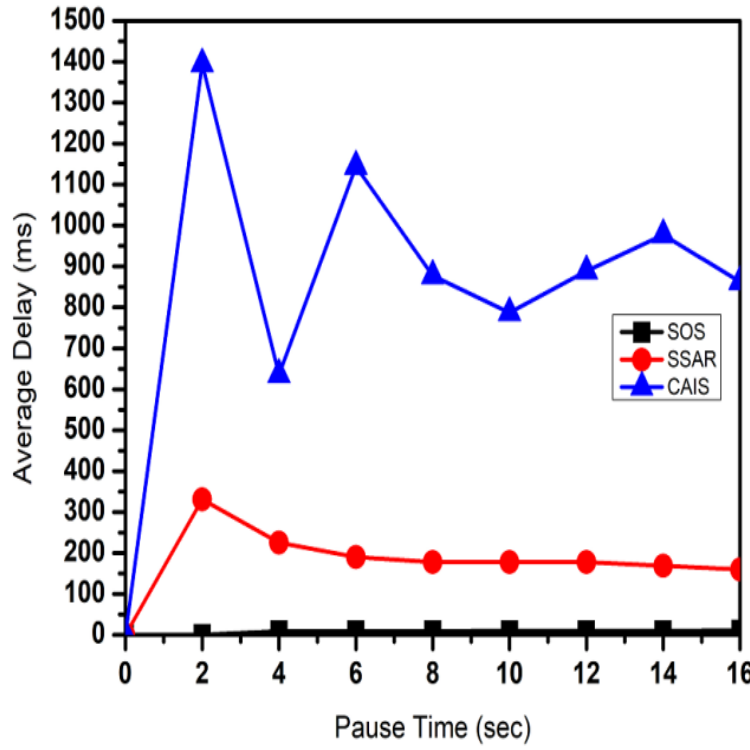}  
  \caption{Average Delivery Delay}
  \label{fig:del25}
\end{subfigure}
\begin{subfigure}{.5\textwidth}
  \centering
  \includegraphics[scale = 1.1]{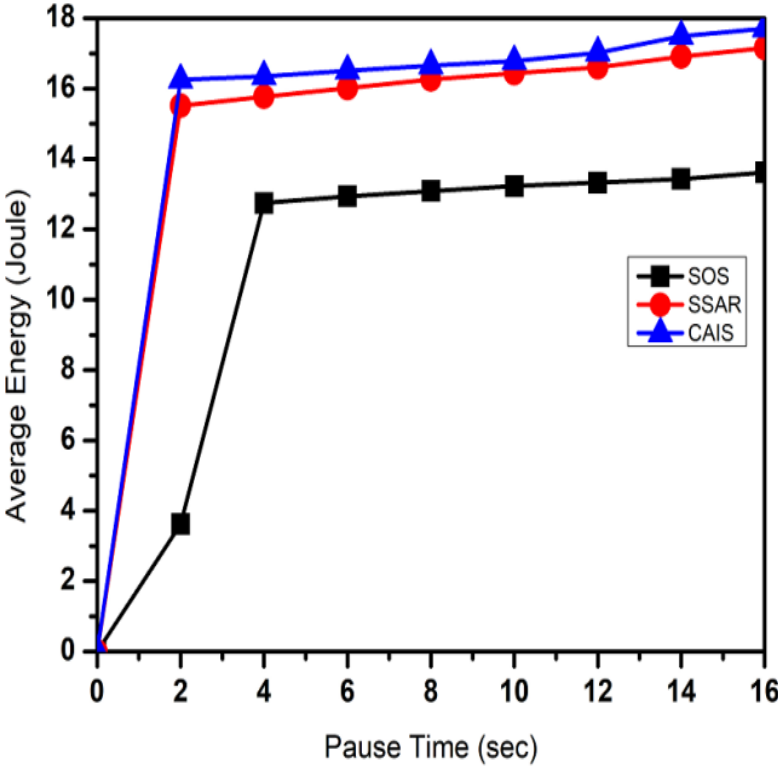}  
  \caption{Average Energy}
  \label{fig:ene25}
\end{subfigure}
\caption{Performance comparisons of the algorithms when 25\% of nodes are selfish}
\label{fig:fig4}
\end{figure}

By injecting 25\% selfish nodes in the network, the packet delivery ratio and throughput of SOS is higher than CAIS and SSAR as showed in Figure~\ref{fig:pkt25} and~\ref{fig:thr25}. 
At pause time 6 sec, the packet delivery ratio of SOS is 21\% and 42\% higher than SSAR and CAIS respectively. This is because of the fact that monitoring nodes constantly monitored the behavior of selfish nodes in SOS. At pause time 8 sec, the throughput of SOS is 36\% and 79\% higher than SSAR and CAIS respectively. The average delay and average energy of SOS is much lower than CAIS and SSAR as shown in Figure~\ref{fig:del25} and~\ref{fig:ene25}. At pause time 4 sec, the average delay of SOS is 22.5 ms that is 14\% lower than SSAR and 41\% lower than CAIS. In addition, the average energy consumed by SOS at pause time 4 sec is 17\% lower than SSAR and 20\% lower than CAIS. Similarly, the average delay and average energy consumed by SOS is lower. Thus, the proposed scheme SOS outperform the both the existing scheme namely SSAR and CAIS in terms of packet delivery ratio, throughput, average delivery delay, and average energy when there are 25\% selfish nodes are present in the network. This is due to the fact that the nodes in the SSAR and CAIS have a weak social relationship with each other and hence ignore to forward messages to other members nodes in a community. The comparison of SOS, SSAR, and CAIS for 25\% selfish nodes for all performance metrics are shown in Table~\ref{tab:comp25}. 
\begin{table}[ht]
\begin{center}
\caption{Performance comparisons of SOS, SSAR, and CAIS for 25\% selfish nodes}
\label{tab:comp25}
\newcolumntype{b}{X}
\newcolumntype{c}{>{\hsize=.1\hsize}X}
\newcolumntype{d}{>{\hsize=.1\hsize}X}
\setlength{\extrarowheight}{2pt}%

\begin{tabularx}{\textwidth}{ c  cccc  d  dd  c  c  d  dd }
\toprule
&&\textbf{PDR}&&& \textbf{Throughput} &&&	\textbf{Avg.Delay}  &&&	\textbf{Avg.Energy}\\
\toprule
\textbf{Pause time}&\textbf{SOS} & \textbf{SSAR } &	\textbf{CAIS}  &	\textbf{SOS} &	\textbf{SSAR} & \textbf{ CAIS} &	\textbf{SOS} &	\textbf{SSAR } &	\textbf{CAIS} &	\textbf{SOS} &	\textbf{SSAR} &	\textbf{CAIS}\\
\midrule

0 &	0.0 &0.0 &0.0& 0.0 &0.0 &0.0 &0.0& 0.0&0.0&	0.0	& 0.0 & 0.0\\
2 &0.4&.68 &.12& 54 & 133.5&6.01&0.0&330.6 &1395& 3.6&15.5&16.2 \\
4 & 0.6 &.60&0.2&209.8& 133.6&16.89& 8.97 & 225.3&635.8&12.7 &	15.7 &16.3\\
6	&0.9&.69&.48&217.6&133.5&26.34&	9.05&190.2&1144&12.9&16.0&16.5\\
8 &	0.9&.60 &.56&220.5 &132.8&28.93&9.17&177.8&877.7&13.0&16.2 & 16.6 \\
10 &0.9 &.42 &.48&220.1 &132.8&26.28&9.62& 177.8&786.9&13.2& 16.4 & 16.7 \\
12 &0.9 &.31 &.49& 220.1 &132.8&26.42&9.62& 177.8&889.0 &	13.3& 16.6 & 17.0\\
14 &0.8&.31&.62	& 220.1&121.9&0.31&9.62& 168.5&977.4 &13.4& 16.9 & 17.4 \\
16 &0.8&.40&.64	& 212.8 &123.4&31.41&10.84& 159.7&862.9&13.6& 17.1 & 17.7 \\
\bottomrule
\end{tabularx}
\end{center}
\end{table}

It can bee seen in Table~\ref{tab:comp25}, at pause time 6 sec, the packet delivery ratio of SOS is 21\% and 42\% higher than SSAR and CAIS respectively. Similarly, at pause time 16 sec, the packet delivery ratio of SOS, SSAR, and CAIS is .8, .40, .64 that is 40\% and 16\% higher than SSAR and CAIS respectively.
This is because of the fact that monitoring nodes constantly monitored the behavior of selfish nodes in SOS. At pause time 8 sec, the throughput of SOS, SSAR, and CAIS is 220.5 kbps, 132.8 kbps, and 28.93 kbps, that is 36\% and 79\% higher than SSAR and CAIS respectively. In addition, At pause time 16 sec, the throughput of SOS, SSAR, and CAIS is 212.8 kbps, 123.4 kbps, and 31.41 kbps. Thus, is still observed that, the throughput of SOS is 37\% and 75\% higher than SSAR and CAIS respectively. Furthermore, the average delay and average energy of SOS is much lower than CAIS and SSAR. At pause time 4 sec, the average delay of SOS is 22.5 ms that is 14\% lower than SSAR and 41\% lower than CAIS.
In addition, At pause time 16 sec, the average delay of SOS, SSAR, and CAIS is 10.84 ms, 159.7 ms, and 862.9 that is 9\% lower than SSAR and 56\% lower than CAIS. Similarly, the average energy consumed by SOS, SSAR, and CAIS at pause time 4 sec is 12.7 joules, 15.7 joules, and 16.3 joules respectively. So it is observed that the energy consumed by SOS is 17\% lower than SSAR and 20\% lower than CAIS. In addition, the average energy consumed by SOS, SSAR, and CAIS at pause time 16 sec is 13.6 joules, 17.1 joules, and 17.7 joules respectively. So it is observed that the energy consumed by SOS is 20\% lower than SSAR and 22\% lower than CAIS. 

\begin{figure}[ht]
\begin{subfigure}{.5\textwidth}
  \centering
  \includegraphics[scale = 1.1]{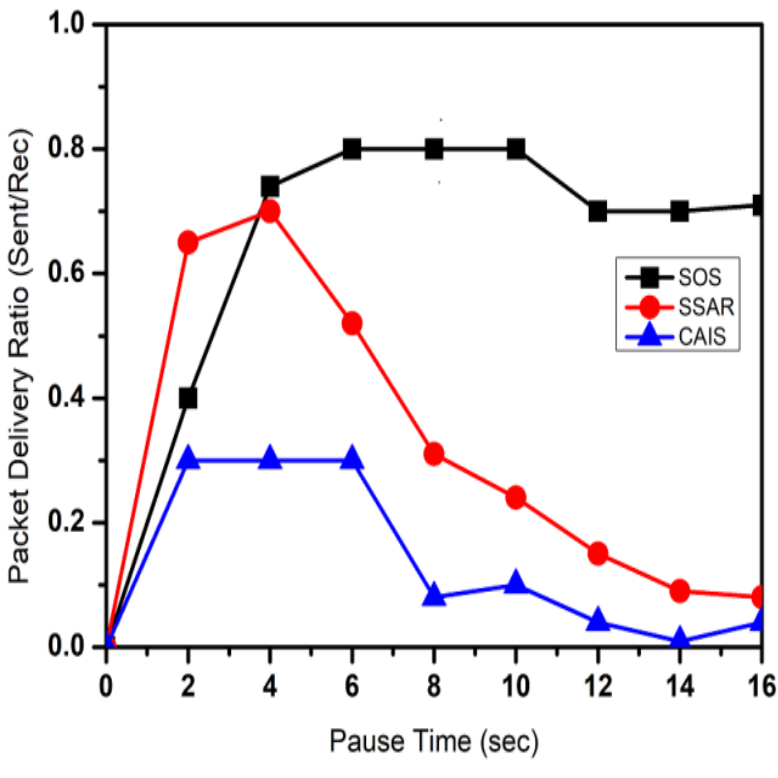}  
  \caption{Packet Delivery Ratio}
  \label{fig:pkt50}
\end{subfigure}
\begin{subfigure}{.5\textwidth}
  \centering
  \includegraphics[scale = 1.1]{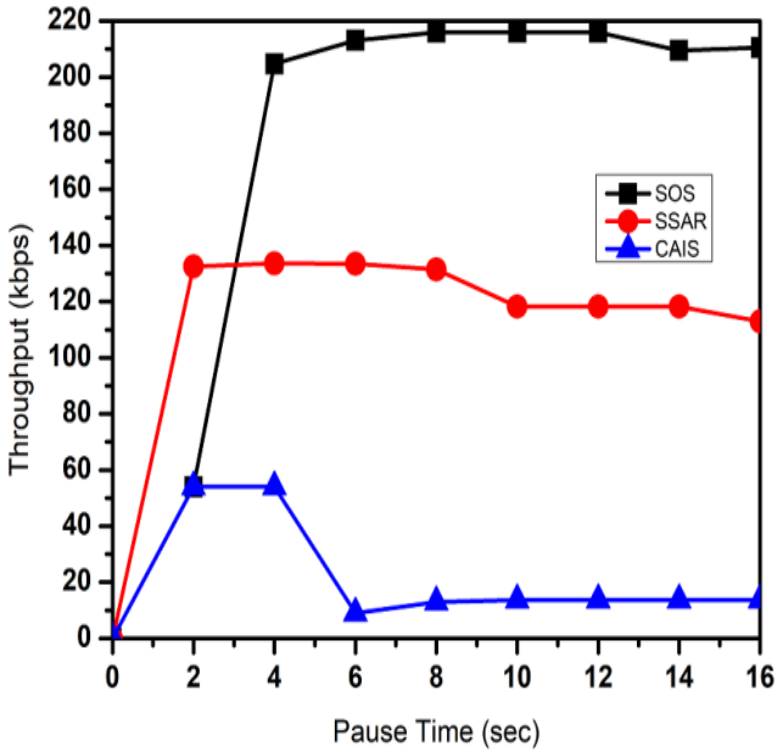}  
  \caption{Throughput}
  \label{fig:thr50} 
\end{subfigure}
\begin{subfigure}{.5\textwidth}
  \centering
  \includegraphics[scale = 1.1]{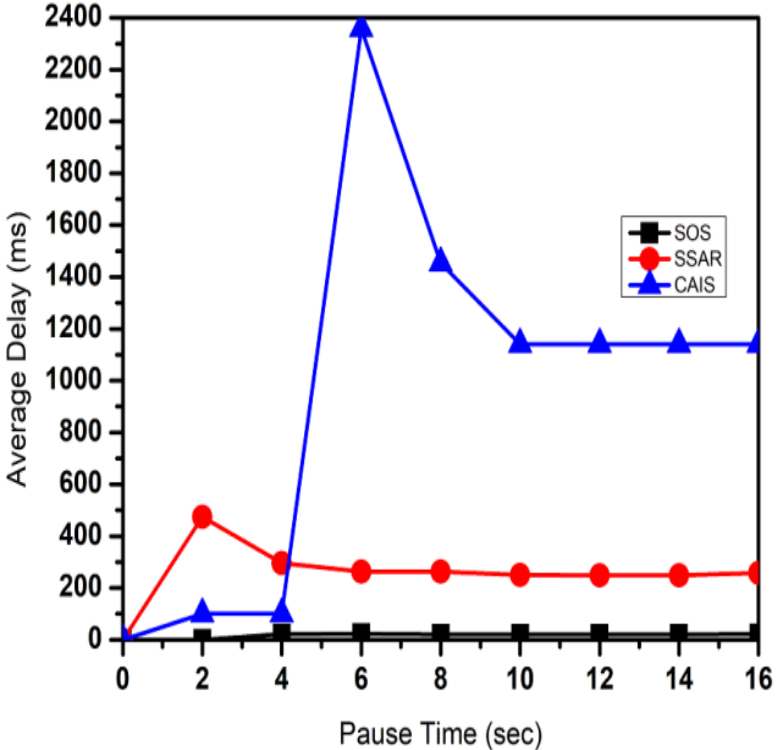}  
  \caption{Average Delivery Delay}
  \label{fig:del50}
\end{subfigure}
\begin{subfigure}{.5\textwidth}
  \centering
  \includegraphics[scale = 1.1]{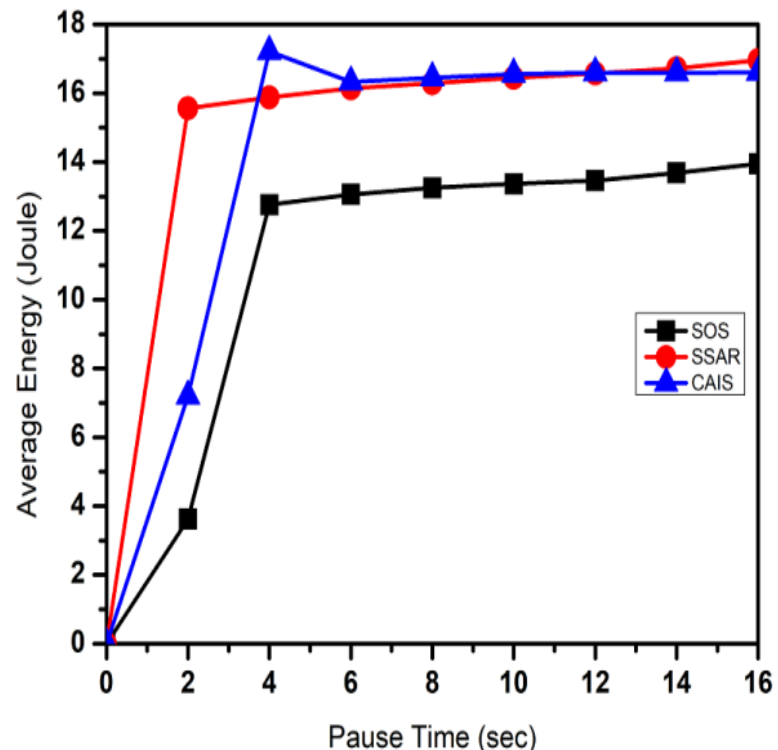}  
  \caption{Average Energy}
  \label{fig:ene50}
\end{subfigure}
\caption{Performance comparisons of the algorithms when 50\% of nodes are selfish}
\label{fig:fig5}
\end{figure}

\begin{figure}[ht]
\begin{subfigure}{.5\textwidth}
  \centering
  \includegraphics[scale = 1.1]{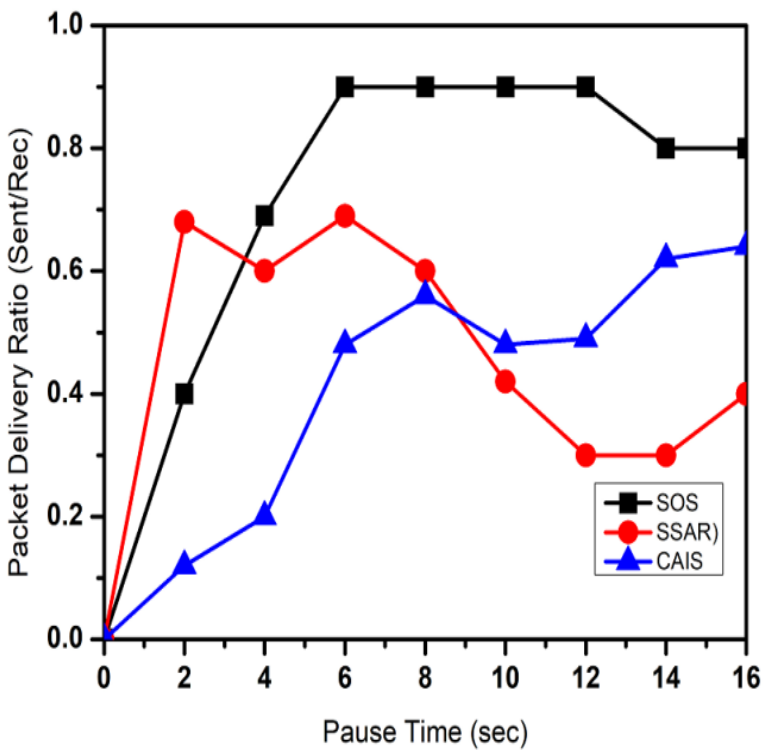}  
  \caption{Packet Delivery Ratio}
  \label{fig:pkt75}
\end{subfigure}
\begin{subfigure}{.5\textwidth}
  \centering
  \includegraphics[scale = 1.1]{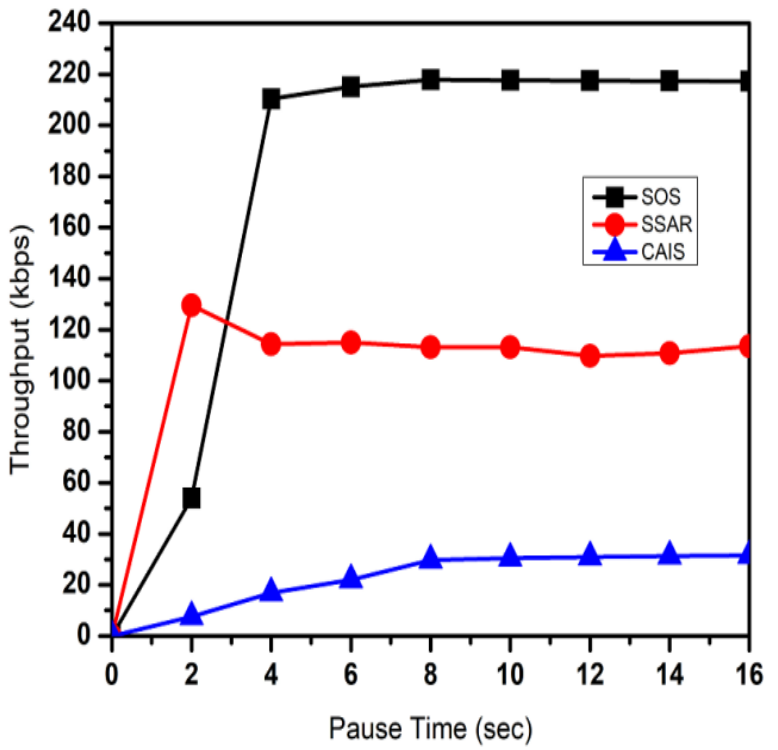}  
  \caption{Throughput}
  \label{fig:thr75} 
\end{subfigure}
\begin{subfigure}{0.5\textwidth}
  \centering
  \includegraphics[scale = 1.1]{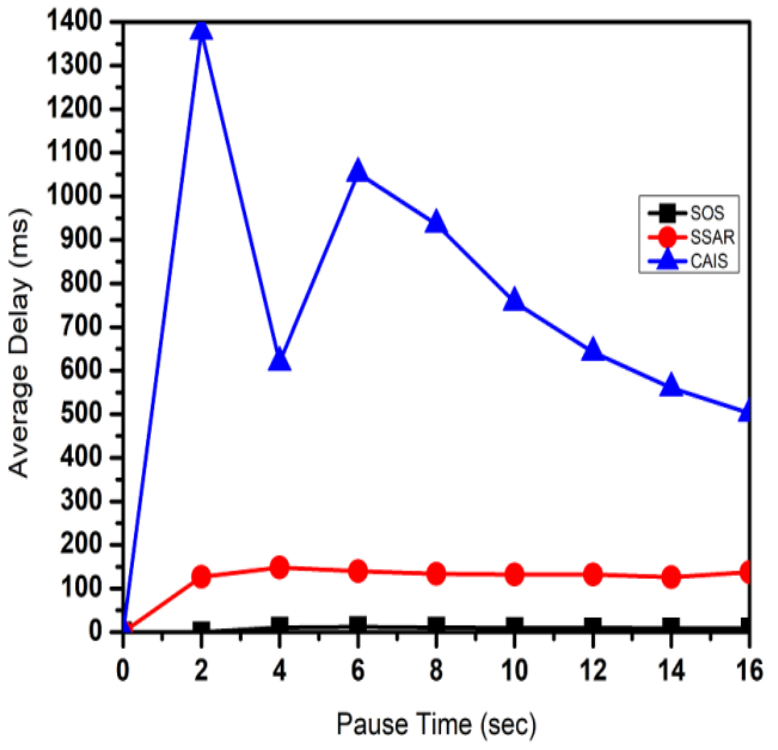}  
  \caption{Average Delivery Delay}
  \label{fig:del75}
\end{subfigure}
\begin{subfigure}{.5\textwidth}
  \centering
  \includegraphics[scale = 1.1]{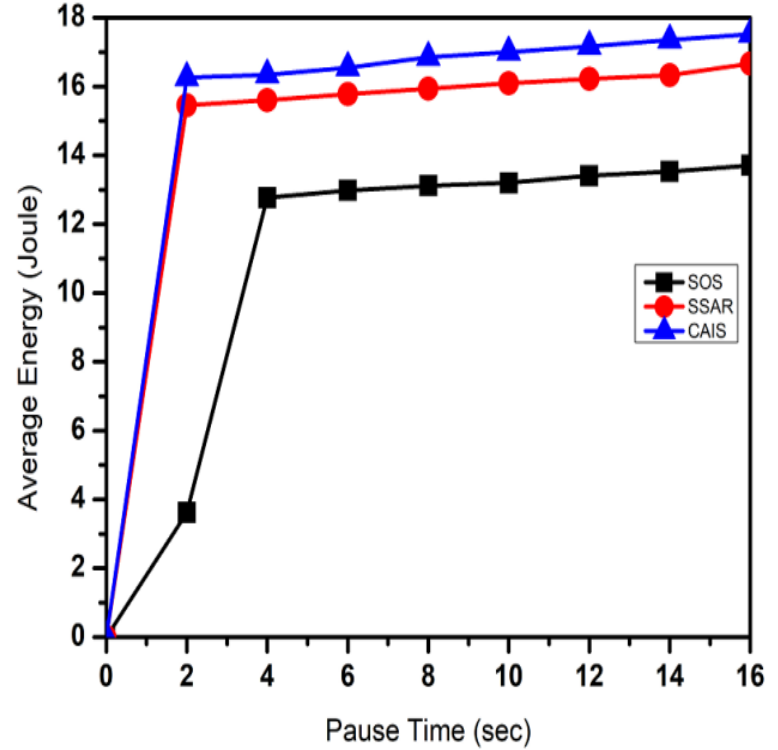}  
  \caption{Average Energy}
  \label{fig:ene75}
\end{subfigure}
\caption{Performance comparisons of the algorithms when 75\% of nodes are selfish}
\label{fig:fig6}
\end{figure}

A similar conclusion can also be drawn by injecting 50\% and 75\% selfish nodes in the simulation as shown in Figure~\ref{fig:fig5} and Figure~\ref{fig:fig6}.

For 50\% selfish nodes in the network, the performance of SOS is better for performance metrics. At pause time 6 sec, the Packets delivery ratio of SOS is .8 (packet sent/rec) that is 28\% and 50\% higher than CAIS and SSAR respectively as shown in Figure~\ref{fig:pkt50}. The throughput of SOS is almost 204.76 kbps at pause time 4 sec that is again 32\% higher than SSAR and 69\% higher than CAIS as shown in Figure~\ref{fig:thr50}. In addition, the average delay of SOS is 20.1 ms at pause time 8 sec that is 59\% lower than CAIS and 10\% lower than SSAR as shown in Figure~\ref{fig:del50}. Similarly, at pause time 4 sec, the average energy consumed by SOS is 12.76 joule that is 18\% lower than SSAR and 15\% lower than CAIS as shown in Figure~\ref{fig:ene50}. The packet delivery ratio of SOS, CAIS, and SSAR at pause time 6 sec are 0.84, 0.35 and 0.09 (packet sent/rec) respectively as shown in Figure~\ref{fig:pkt75}. Thus, the packet delivery ratio of SOS is 49\% and 75\% higher than CAIS and SSAR respectively. At pause time 8 sec, the throughput of SOS is 217.91 kbps that is 43\% and 78\% higher than SSAR and CAIS respectively as shown in Figure~\ref{fig:thr75}. In addition, the average delay of SOS at pause time 4 sec is 10.7 ms that is 9\% lower than SSAR and 43\% lower than CAIS as shown in Figure~\ref{fig:del75}. Similarly, at pause time 4 sec, again the average energy consumed by SOS is 12.77 joule that is 17\% and 20\% lower than SSAR and CAIS respectively as shown in Figure~\ref{fig:ene75}. 

\begin{figure}[ht]
\begin{subfigure}{.5\textwidth}
  \centering
  \includegraphics[scale = 0.51]{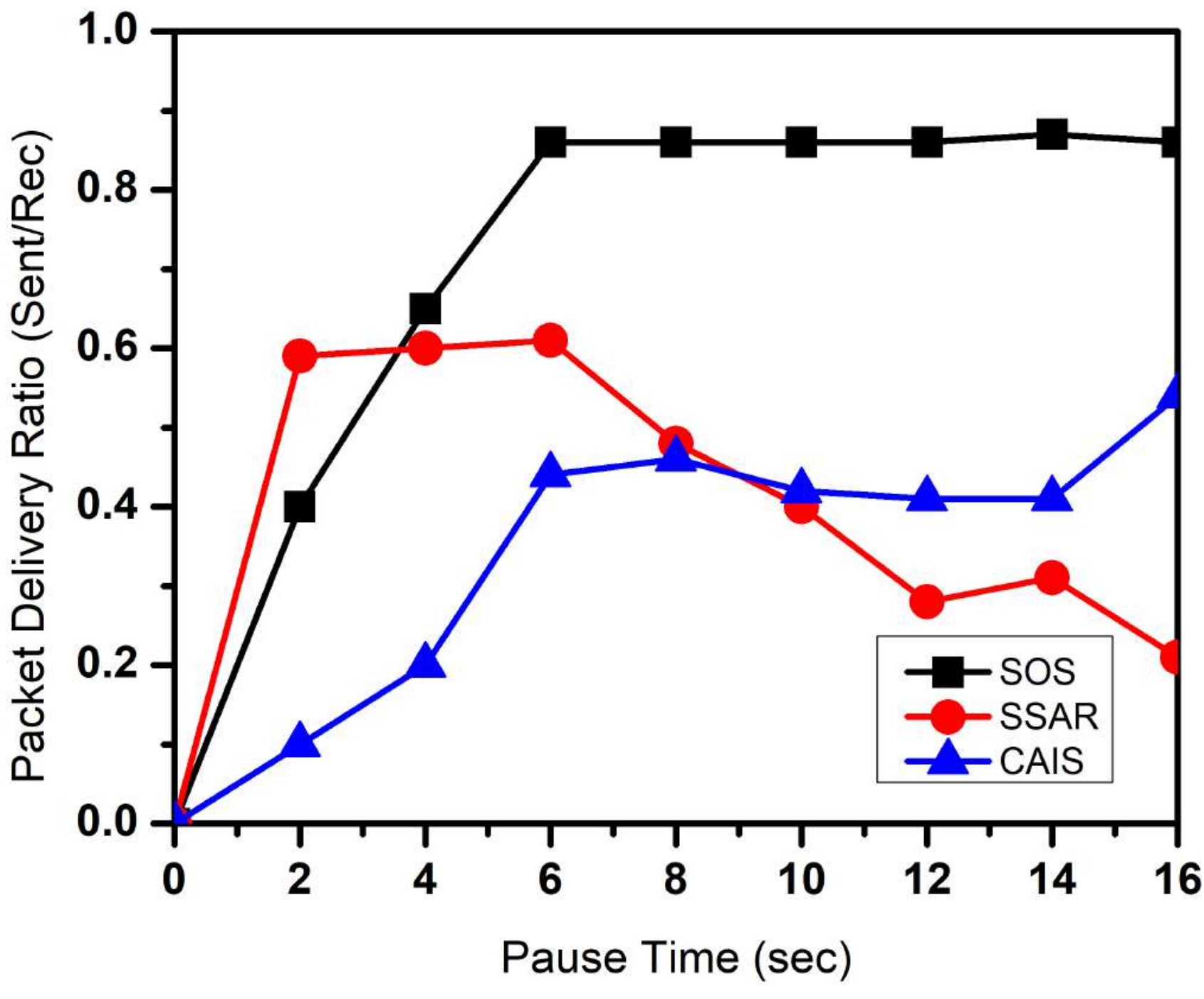}  \caption{Packet Delivery Ratio}
  \label{fig:pkt90}
\end{subfigure}
\begin{subfigure}{.5\textwidth}
  \centering
  \includegraphics[scale = 0.51]{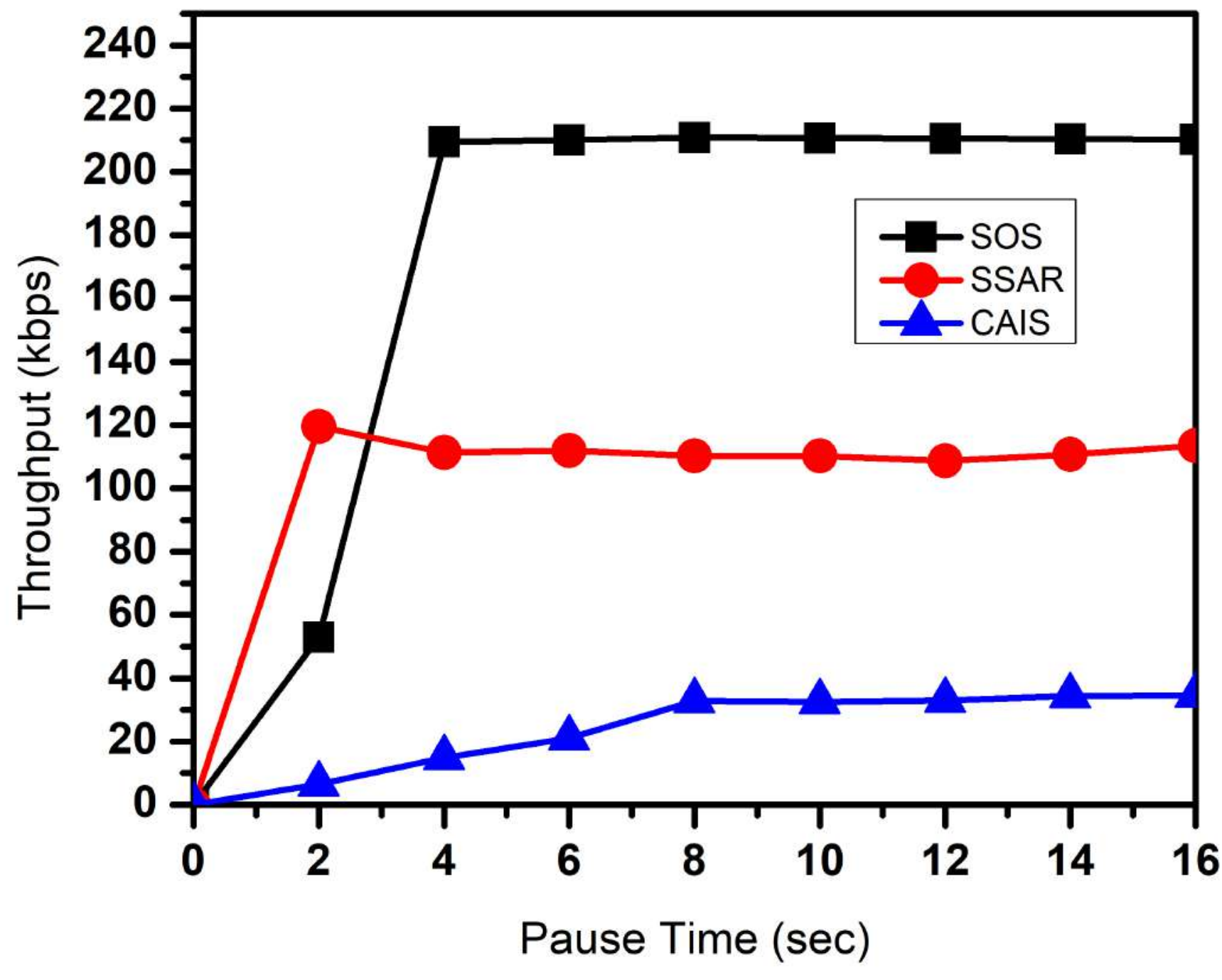}  
  \caption{Throughput}
  \label{fig:thr90} 
\end{subfigure}
\begin{subfigure}{.5\textwidth}
  \centering
  \includegraphics[scale = 0.51]{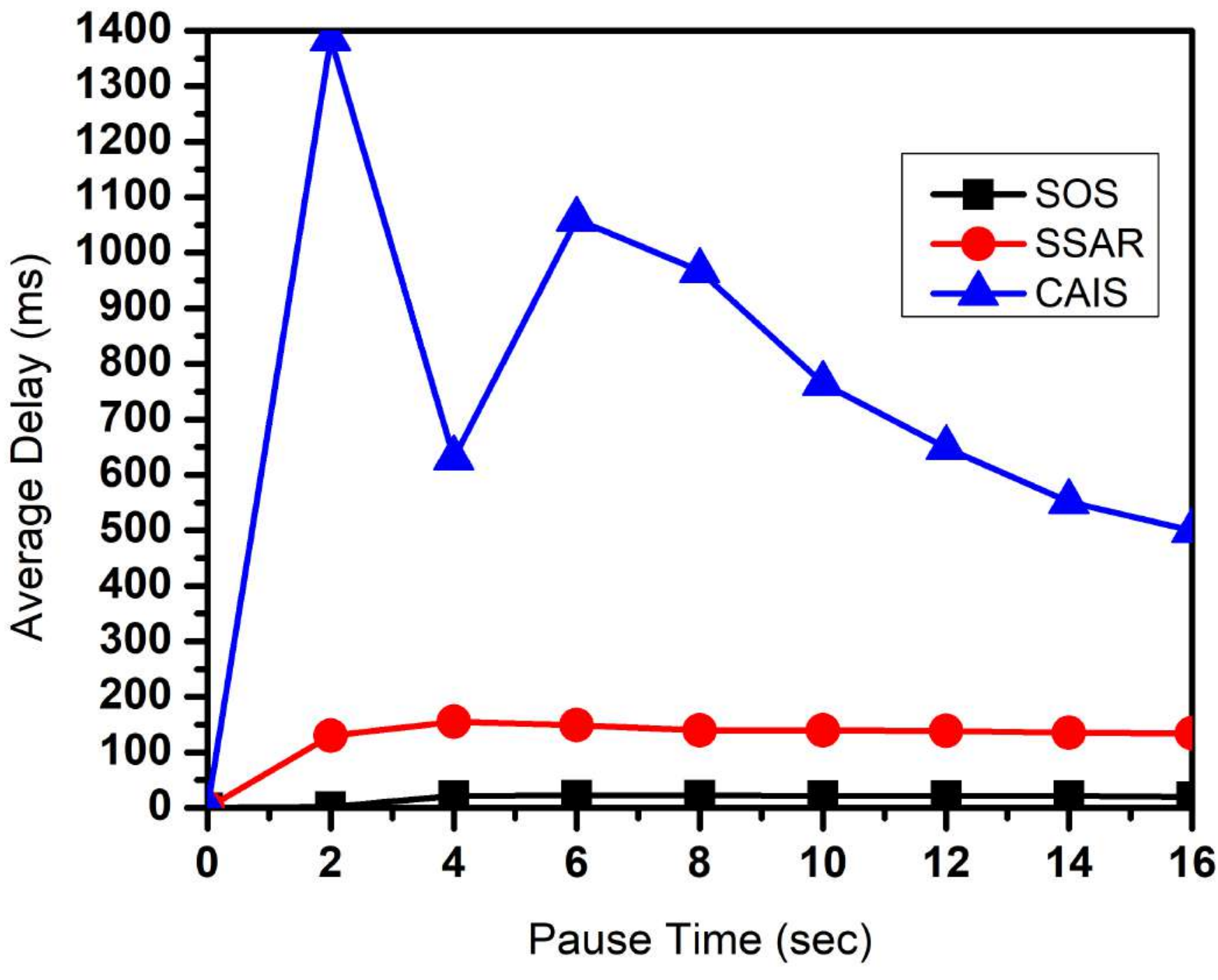}  
  \caption{Average Delivery Delay}
  \label{fig:del90}
\end{subfigure}
\begin{subfigure}{.5\textwidth}
  \centering
  \includegraphics[scale = 0.52]{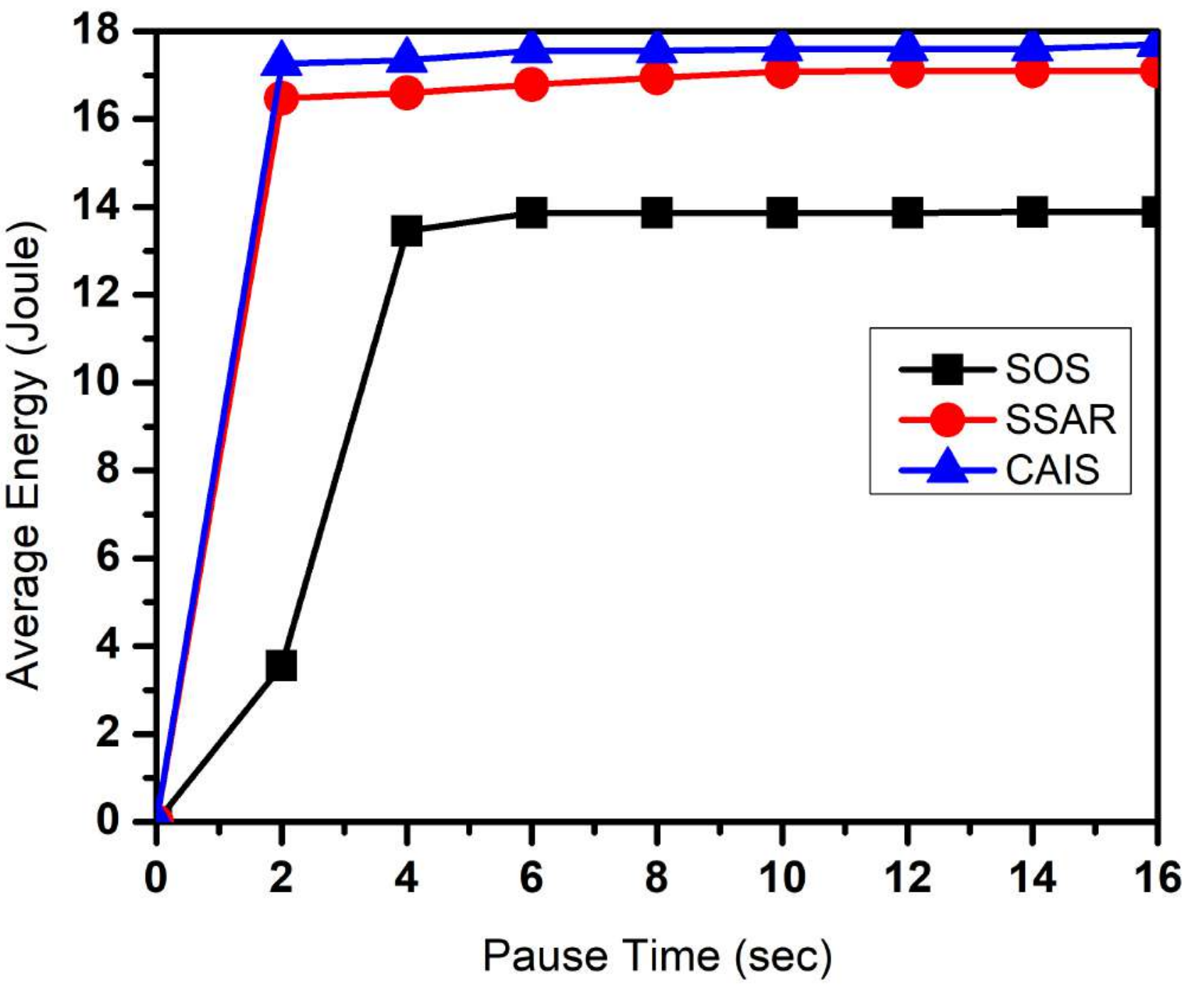}  
  \caption{Average Energy}
  \label{fig:ene90}
\end{subfigure}
\caption{Performance comparisons of the algorithms when 90\% of nodes are selfish}
\label{fig:fig7}
\end{figure}

By injecting 90\% selfish nodes in the network, the SOS scheme still outperforms SSAR and CAIS in terms of packet delivery ratio, throughput, average delay, and average energy as shown in Figure~\ref{fig:fig7}.
It can be seen in Figure~\ref{fig:pkt90}, at pause time 8 sec, the packet delivery ratio of SOS, SSAR, and CAIS are .86, .48, and .46 that is 38\% and 40\% higher than SSAR and CAIS respectively. In addition, at pause time 14 sec, the packet delivery ratio of SOS, SSAR, and CAIS are .87, .31, and .41 that is 56\% and 46\% higher than SSAR and CAIS respectively. Similarly, at pause time 8 sec, the throughput of the SOS, SSAR, and CAIS scheme are 210.91 kbps, 110.14 kbps, and 32.76 kbps, that is 42\% and 74\% higher than SSAR and CAIS respectively as shown in Figure~\ref{fig:thr90}. In addition, at pause time 14 sec, the throughput of the SOS, SSAR, and CAIS scheme are 210.37 kbps, 110.73 kbps, and 34.29 kbps, that is still 41\% and 73\% higher than SSAR and CAIS respectively. Furthermore, at pause time 8 sec, the average delay of SOS, SSAR, and CAIS are 22.72 ms, 139.89 ms, and 967.83 ms, that is 7.3\% and 67\% lower than SSAR and CAIS respectively as shown in Figure~\ref{fig:del90}. In addition, at pause time 14 sec, the average delay of SOS, SSAR, and CAIS are 21.27 ms, 134.95 ms, and 551.15 ms, that is still 8\% and 38\% lower than SSAR and CAIS respectively.

Similarly, at pause time 8 sec, the average energy consumed by SOS, SSAR, and CAIS are 13.86 joules, 16.94 joules, and 17.55 joules, that is 17\% and 20\% lower than SSAR and CAIS respectively as shown in Figure~\ref{fig:ene90}. Similarly, at pause time 16 sec, the average energy of SOS is 13.89 joules that is 18\% and 21\% lower than SSAR and CAIS respectively. The comparison of SOS, SSAR, and CAIS for 90\% selfish nodes for all performance metrics are shown in Table~\ref{tab:comp90}. 

\begin{table}[ht]
\begin{center}
\caption{Performance comparisons of SOS, SSAR, and CAIS for 90\% selfish nodes}
\label{tab:comp90}
\newcolumntype{b}{X}
\newcolumntype{c}{>{\hsize=.1\hsize}X}
\newcolumntype{d}{>{\hsize=.1\hsize}X}
\setlength{\extrarowheight}{2pt}%

\begin{tabularx}{\textwidth}{ c  cccc  d  dd  c  c  d  dd }
\toprule
&&\textbf{PDR}&&& \textbf{Throughput} &&&	\textbf{Avg.Delay}  &&&	\textbf{Avg.Energy}\\
\toprule
\textbf{Pause time}&\textbf{SOS} & \textbf{SSAR } &	\textbf{CAIS}  &	\textbf{SOS} &	\textbf{SSAR} & \textbf{ CAIS} &	\textbf{SOS} &	\textbf{SSAR } &	\textbf{CAIS} &	\textbf{SOS} &	\textbf{SSAR} &	\textbf{CAIS}\\
\midrule

0 &	0.0 &0.0 &0.0& 0.0 &0.0 &0.0 &0.0& 0.0&0.0&	0.0	& 0.0 & 0.0\\
2 &0.4&.59 &.10& 53 & 119.46&6.52&2&129.82&1385.62& 3.55&16.47&17.26 \\
4 & .65 &.60&.20&209.4& 111.25&14.9& 21.75 & 155.0& 630.54&13.45 &	16.60 &17.34\\
6&.86&.61&.44&210.12&111.93&21.03&	22.28&149.14&1060.18&13.86&16.78&17.55\\
8 &	.86 &.48&.46 &210.91&110.14&32.76&22.72&139.89&967.83&13.86&16.94 & 17.55 \\
10 &.86 &.40 &.42&210.66 &110.16&32.47&21.98& 139.19&764.24&13.86& 17.09 & 17.60 \\
12 &.86 &.28&.41& 210.49 &108.71&32.94&21.7& 139.86&648.28 &13.86& 17.10 & 17.60\\
14 &.87&.31&.41	& 210.37&110.73&34.29&21.27& 134.95&551.15 &13.89& 17.10 & 17.60 \\
16 &.86&.21&.54	& 210.28 &113.45&34.54&20.14& 134.67&499.33&13.89& 17.10 & 17.70 \\
\bottomrule
\end{tabularx}
\end{center}
\end{table}

It can be seen in Table~\ref{tab:comp90}, the packet delivery ratio of SOS, SSAR, and CAIS are .86, .48, and .46 that is 38\% and 40\% higher than SSAR and CAIS respectively. In addition, at pause time 16 sec, the packet delivery ratio of SOS is .86 that is 65\% and 32\% higher than SSAR and CAIS respectively. Similarly, at pause time 8 sec, the throughput of the SOS, SSAR, and CAIS scheme are 210.91 kbps, 110.14 kbps, and 32.76 kbps, that is 42\% and 74\% higher than SSAR and CAIS respectively. In addition, at pause time 16 sec, the throughput of the SOS, SSAR, and CAIS scheme are 210.28 kbps, 113.45 kbps, and  34.54 kbps, that is 38\% and 73\% higher than SSAR and CAIS respectively. In addition, at pause time 8 sec, the average delay of SOS, SSAR, and CAIS are 22.72 ms, 139.89 ms, and 967.83 ms, that is 7.3\% and 67\% lower than SSAR and CAIS respectively. Furthermore, at pause time 16 sec, the average delay of SOS, SSAR, and CAIS are 20.14 ms, 134.67 ms, and 499.33 ms, that is 8\% and 34\% lower than SSAR and CAIS respectively.

Similarly, at pause time 8 sec, the average energy consumed by SOS, SSAR, and CAIS are 13.86 joules, 16.94 joules, and 17.55 joules, that is 17\% and 20\% lower than SSAR and CAIS respectively. Similarly, at pause time 16 sec, the average energy of SOS is 13.89 joules that is 18\% and 21\% lower than SSAR and CAIS respectively. It is due to the fact that it stimulates the nodes to participate in the network and forward messages in a cooperative manner in its community. Cooperative nodes are given some incentive in the form of reputation. The nodes showing selfish behavior are penalized in the form of expulsion from the network. However, nodes are not directly expelled from the network, it has given a warning first. The two other techniques have not considered the effect of the selfish nodes on the community. Thus, the results demonstrated in SOS shows that it has encouraged the nodes in a community to cooperate with all other nodes in message delivery. Therefore, the SOS scheme outperformed the existing two techniques namely CAIS and SSAR in terms of packet delivery ratio, throughput, average delivery delay, and average energy consumed.

\section{CONCLUSION AND FUTURE WORK}
\label{con}
In this article, a new SOS scheme is proposed to stimulate the selfish nodes in different communities to cooperatively forward messages for other nodes. The proposed approach is based on the electoral system and generally omits selfishness in IoT based SCC. In an electoral system, different heads are elected such as Community Head, Incentive Head and monitoring Head in the communities. These heads are elected based on two characteristics weight and cooperation. Incentive in the form of reputation is awarded to the nodes for their cooperation within the community using VCG model. Furthermore, nodes are also penalized for showing repeated selfish behavior. In the proposed scheme SOS, one of the important rules called the Collective Importance Factor (CIF) principle is used that decides the selfish and cooperative nature of nodes. This rule computes the trust depends on evidence from distinct nodes. These evidences are then merged by using the Extended Dempster-Shafer model to resolve uncertainty situations. For comparative analysis, two protocols namely SSAR and CAIS are thoroughly simulated and analyzed. The results in terms of data delivery ratio, network delay and average energy consumed are compared with the proposed approach. The results indicate that SOS can possibly accommodate a large number of selfish nodes by enabling them to collaborate in a community to improve network performance. As a future enhancement of the proposed approach, the bandwidth may be considered as reputation criteria of nodes in the network for service delivery.


\bibliography{wileyNJD-AMA}%
\end{document}